\documentclass[twocolumn,pre,nofootinbib,floatfix,superscriptaddress]{revtex4}

\usepackage{dcolumn}
\usepackage{graphicx}
\usepackage{rotating}
\usepackage{amsmath,amsfonts,amssymb}
\usepackage{subfigure}

\newcommand{\remove}[1]{}

\begin{document}

\title{The performance of modularity maximization in practical contexts}
\author{Benjamin H. Good}
\email{conkerll@gmail.com}
\affiliation{Department of Physics, Swarthmore College, Swarthmore PA, 19081 USA}
\affiliation{Santa Fe Institute, 1399 Hyde Park Road, Santa Fe NM, 87501, USA}
\author{Yves-Alexandre de Montjoye}
\email{YvesAlexandre@deMontjoye.com}
\affiliation{Department of Applied Mathematics, Universit\'e Catholique de Louvain, 4 Avenue Georges Lemaitre, B-1348 Louvain-la-Neuve, Belgium}
\affiliation{Santa Fe Institute, 1399 Hyde Park Road, Santa Fe NM, 87501, USA}
\author{Aaron Clauset}
\email{aaronc@santafe.edu}
\affiliation{Santa Fe Institute, 1399 Hyde Park Road, Santa Fe NM, 87501, USA}

\begin{abstract}
Although widely used in practice, the behavior and accuracy of the popular module identification technique called {\em modularity maximization} is not well understood in practical contexts. Here, we present a broad characterization of its performance in such situations. First, we revisit and clarify the resolution limit phenomenon for modularity maximization. Second, we show that the modularity function $Q$ exhibits extreme degeneracies: it typically admits an exponential number of distinct high-scoring solutions and typically lacks a clear global maximum. Third, we derive the limiting behavior of the maximum modularity $Q_{\max}$ for one model of infinitely modular networks, showing that it depends strongly both on the size of the network and on the number of modules it contains. Finally, using three real-world metabolic networks as examples, we show that the degenerate solutions can fundamentally disagree on many, but not all, partition properties such as the composition of the largest modules and the distribution of module sizes. These results imply that the output of any modularity maximization procedure should be interpreted cautiously in scientific contexts. They also explain why many heuristics are often successful at finding high-scoring partitions in practice and why different heuristics can disagree on the modular structure of the same network. We conclude by discussing avenues for mitigating some of these behaviors, such as combining information from many degenerate solutions or using generative models.
\end{abstract} 

\maketitle

\section{Introduction}
Networks are a powerful tool for understanding the structure, dynamics, robustness and evolution of complex biological, technological and social systems~\cite{wasserman:faust:1994,newman:2003}. The automatic detection of modular structures in networks---also called communities~\cite{newman:girvan:2004} or compartments~\cite{allesina:pascual:2009}, and conventionally understood to be large subgraphs with high internal densities---can provide a scalable way to identify functionally important or closely related classes of nodes from interaction data alone~\cite{porter:onnela:mucha:2009,fortunato:2010}.

The identification of modular structures has broad implications for many systems-level questions. For instance, it has strong consequences for the behavior of dynamical processes on networks~\cite{arenas:etal:2006,restrepo:ott:hunt:2006}, and can provide a principled way to reduce or coarse-grain a system by dividing global heterogeneity into relatively homogeneous substructures. The search for such modular substructures has been particularly intense in molecular networks. This is, in part, because modules have theoretical significance for molecular networks~\cite{hartwell:etal:1999,barabasi:oltvai:2004,papin:reed:palsson:2004}: they can correspond to functional clusters of genes or proteins~\cite{huss:holme:2007,singh:etal:2008}, they may represent targets of natural selection above the level of individual genes or proteins but below the level of the whole organism, and they may provide evidence of past evolutionary constraints or pressures~\cite{singh:etal:2008,guimera:amaral:2005}. Past work along these lines has identified modular structures in signaling, metabolic and protein-interaction systems~\cite{guimera:amaral:2005,rives:galitski:2003,spirin:mirny:2003,zhao:etal:2007}, although some questions remain about their statistical significance~\cite{karrer:etal:2008} and functional relevance~\cite{pinkert:schultz:reichardt:2008}. Naturally, many of these questions apply equally well to modules in social and technological networks.

Empirical evidence for a modular organization is typically derived using computer algorithms that automatically identify modules using network connectivity data, and among the many techniques now available (see Refs.~\cite{porter:onnela:mucha:2009,fortunato:2010,newman:2004:a} for reviews), the method of {\em modularity maximization}~\cite{newman:girvan:2004} is by far the most popular. Under this method, each decomposition or partition of a network into $k$ disjoint modules is given a score $Q$, called the {\em modularity} or sometimes ``Newman-Girvan modularity'':
\begin{align}
\label{eq:modularity}
Q & = \sum_{i=1}^{k} \left[  \frac{e_i}{m} - \left( \frac{d_i}{2m} \right) ^2 \right] \enspace ,
\end{align}
where $e_i$ is the number of edges in module $i$, $d_i$ is the total degree of nodes in module $i$ and $m$ is the total number of edges in the network.  Intuitively, $Q$ measures the difference between the observed connectivity within modules and its expected value for a random graph with the same degree sequence~\cite{molloy:reed:1995}. Thus, a ``good'' partition---with $Q$ closer to unity---identifies groups with many more internal connections than expected at random; in contrast, a ``bad'' partition---with $Q$ closer to zero---identifies groups with no more internal connections than we expect at random. This reasonable formulation recasts the problem of identifying modules as a problem of finding the so-called \emph{optimal partition}, i.e., the partition that maximizes the modularity function $Q$.

Despite the popularity of modularity maximization, much remains unknown about the quality and significance of its output when applied to real-world networks with unknown modular structure. Most past work has focused on developing new ways of detecting modules, rather than on characterizing their performance in practical situations. In general, maximizing $Q$ is known to be NP-hard~\cite{brandes:etal:2008}, but many heuristic approaches---including greedy agglomeration~\cite{newman:2004:b,clauset:etal:2004,blondel:etal:2008}, mathematical programming~\cite{agarwal:kempe:2008}, spectral methods~\cite{newman:2006,richardson:mucha:porter:2009}, extremal optimization~\cite{duch:arenas:2005}, simulated annealing~\cite{guimera:amaral:2005} and sampling techniques~\cite{massen:doye:2006,sales-pardo:etal:2007}---perform well on simple synthetic networks with strong modular structure~\cite{newman:girvan:2004} and they often succeed at finding high-modularity partitions in practice. The apparent success of these methods has led to their widespread adoption, and often (but not always~\cite{sales-pardo:etal:2007,fortunato:barthelemy:2007}) the implicit acceptance of several assumptions: (i) empirical networks with modular structure tend to exhibit a clear optimal partition~\cite{karrer:etal:2008,massen:doye:2006}, (ii) high-modularity partitions of an empirical network are structurally similar to this optimal partition and (iii) the estimated $Q_{\max}$ can be meaningfully compared across networks~\cite{kreimer:etal:2008}.

Here, we present a broad characterization of the behavior of modularity maximization in practical contexts. First, we revisit and clarify the resolution limit phenomenon~\cite{fortunato:barthelemy:2007,kumpula:etal:2007,branting:2008,berry:etal:2009}. We then show that the above assumptions do not hold when modularity maximization is applied to networks with modular or hierarchical structure. Using a combination of analytic and numerical techniques (whose implementations are available online\footnote{See \texttt{http://www.santafe.edu/\~{}aaronc/modularity/}}), we show that the modularity function $Q$ exhibits extreme degeneracies such that the globally maximum modularity partition is typically hidden among an exponential number of structurally dissimilar, high-modularity solutions. We then derive the asymptotic behavior of $Q_{\max}$ in the limit of infinitely modular networks, showing that it depends strongly on both the size of the network and on the number of modules it contains. Finally, using three real-world examples of metabolic networks, we show that the degenerate solutions can fundamentally disagree on many (but not all) partition properties such as the composition of the largest identified modules and the distribution of module sizes. This latter finding poses a serious problem for scientific applications.

Together, these results significantly extend our understanding of the behavior and results of modularity maximization in practical contexts. When applied to networks with modular structure, these results imply that any particular partition of a real-world network found by maximizing modularity should be interpreted cautiously. In principle, there is nothing special about the modularity function with respect to the degeneracy result and we expect that other module identification techniques will exhibit similar behavior. We conclude by discussing the prospects of ameliorating these issues, for instance, by combining information across degenerate solutions or using generative models.


\section{The Resolution Limit Revisited}
\label{section:resolution:limit}

Recently, Fortunato and Barth\'{e}lemy showed that modularity admits an implicit resolution limit~\cite{fortunato:barthelemy:2007}, in which the maximum modularity partition can fail to resolve modules smaller than a size, or weight~\cite{berry:etal:2009}, that depends on the total weight of edges in the network $m$. This violates the notion that the quality of a module should be a local property.
As a result, intuitively modular structures, such as cliques of moderate size, can be hidden within large agglomerations that 
yield a higher modularity score, and the peak of the modularity function (the optimal partition) may not coincide with the partition that correctly identifies these intuitive modules (the intuitive partition). 

This behavior is sometimes described as an implicit preference on edge weight within identified modules. But as we show here, the resolution limit is better understood as a consequence of assuming that inter-module connections follow a random graph model, which induces an explicit preference on the weight of between-module connections.
Thus, it should not be surprising that other partition score functions that make similar random-graph assumptions about inter-module edges, such as Potts-models~\cite{kumpula:etal:2007} and several likelihood-based~\cite{branting:2008} techniques, also exhibit resolution limits.



To begin, we consider the change in modularity $\Delta Q$ obtained by merging two modules in the intuitive partition. If this change is positive, then the modularity function will fail to resolve the intuitive modules, since a higher modularity score is achieved by merging them. Let $e_i$ and $e_j$ be the number of edges within the modules and $e_{ij}$ be the number of edges between them.  The change in $Q$ for merging them~\cite{fortunato:barthelemy:2007} is 
\begin{align}
\label{eq:merge-penalty}
\Delta Q_{ij} = \frac{e_{ij}}{m} - 2 \left( \frac{d_i}{2m} \right) \left( \frac{d_j}{2m} \right) \enspace ,
\end{align}
which is positive whenever
\begin{align}
\label{eq:merge:requirement}
e_{ij} & > \frac{d_i d_j}{2m} \enspace .
\end{align} 
That is, independent of the modules' internal structure, a merger is beneficial whenever the observed number of inter-module edges $e_{ij}$ exceeds the number expected for a random graph with the same degree sequence $\langle e_{ij} \rangle=d_{i}d_{j}/2m$~\cite{fortunato:2010}. This behavior is particularly problematic for large unweighted networks because modularity tends to expect $\langle e_{ij} \rangle < 1$ while the minimum inter-module weight is $e_{ij}=1$, i.e., a single edge. On the other hand, as we show below, weighted networks whose inter-module connectivity approximates the null expectation can escape the resolution limit completely.

\subsection{Two examples}

\begin{figure}[t]
\begin{center}
\begin{tabular}{c}
\includegraphics[scale=0.9]{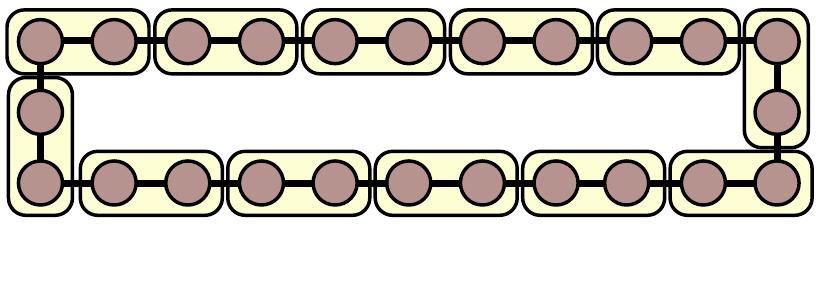}
\end{tabular}
\caption{(color online) A schematic of a ring network with $k=24$ cliques of $c=5$ nodes each (shaded circles) joined by single links to form a ring. The intuitive partition, which places individual cliques on their own, has modularity $Q_{1}=0.8674$, while the optimal partition (the 2-clique tiles), which merges adjacent cliques, has slightly larger modularity $Q_{2}=0.8712$.  }
\label{fig:ring}
\end{center}
\end{figure}

To illustrate the subtlety of this behavior, we consider two versions of the ring network model~\cite{fortunato:barthelemy:2007}, in which $k$ cliques, each containing $c$ nodes, are connected by single edges to form a ring (Fig.~\ref{fig:ring}). The intuitive partition here places each clique in a group by itself and, at least for small values of $k$, this is also the optimal partition.  The penalty for merging a pair of adjacent cliques is given by
\begin{equation}
\label{eq:ring-merge-penalty}
\Delta Q = \frac{1}{k \left[ {c \choose 2} + 1 \right]} - 2k^{-2} \enspace ,
\end{equation}
which is positive whenever
\begin{equation}
\label{eq:resolution-limit-condition}
k > 2 {c \choose 2} + 2 \enspace .
\end{equation}
Thus, there is some number of cliques $k^{*}$ above which the modularity function gives a higher score to a partition that merges pairs of adjacent cliques.

[We note that this argument generalizes to merging $\ell$ adjacent cliques: an \mbox{$\ell$-merged} partition has greater modularity than an \mbox{$(\ell-1)$-merged} partition whenever
\begin{align}
k & > \ell(\ell-1) \left[ {c \choose 2} + 1\right] \enspace .
\end{align}
Thus, the resolution limit is multi-scale and for large connected networks, intuitively modular structures can be hidden within very large agglomerations.]

The resolution limit appears in the ring network because each module is connected to its nearest neighbors with a constant weight $e_{st} = O(1)$, while the null model expects this weight to decrease with $k$. Thus, there must be some size of the network, i.e., a value of $k$, above which even a single unweighted edge between two cliques will appear ``surprising'' under the null model, and modularity will favor merging these minimally connected modules.




Some kinds of weighted networks, however, do not exhibit a resolution limit. For instance, consider a weighted variation of the ring network, 
composed of $k$ cliques, each containing $c$ nodes and each with internal weight \mbox{$e_i = {c \choose 2}$}. Now, instead of connecting each clique to two others to form a ring, we take the same total weight and spread it evenly across connections to every other clique. That is, we completely connect the cliques using edges with weight $e_{ij} = 2/(k-1)$. Notably, the total weight of a module here is exactly the same as in the example above, that is, $d_{i} = {c \choose 2} + 2$, and the total weight in the network grows with $k$. But, the penalty for merging a pair of connected cliques in this network is given by
\begin{equation}
\Delta Q = \frac{2}{k(k-1) \left[ {c \choose 2} + 1 \right]} - 2k^{-2} \enspace ,
\end{equation}
which is negative for all $k > 2$. Thus, it is never beneficial to merge a pair of cliques and there is no resolution limit in this network.

Surprisingly, despite the fact that the internal and external module weights in both of these toy networks are exactly the same, one exhibits a resolution limit while the other does not. The crucial difference between these examples is that in one the weight of an inter-module connection is independent of the size of the network, while in the other, it decreases. This dependence allows the observed inter-module connectivity between any pair of modules to closely follow the connectivity expected under the null model, and to avoid the condition given by Eq.~\eqref{eq:merge:requirement} for merging two modules, even though the total weight in the network grows without bound.
%

Thus, we see that the resolution limit is better explained as a systematic deviation between the inter-module connectivity $e_{ij}$ and the connectivity expected under the random-graph null model $\langle e_{ij} \rangle=d_{i}d_{i}/2m$, than as an implicit preference on the weight of edges within modules.

\subsection{A broader perspective}

For unweighted networks, and for many weighted ones, Fortunato and Barth\'{e}lemy correctly argue that the resolution limit poses a very real problem for the direct interpretation of the optimal partition's composition.

On the other hand, since the intuitive partition is always a refinement of the optimal partition, cleverly designed algorithms may be able to circumvent the resolution limit in some cases. For instance, divisive algorithms that recursively partition large modules~\cite{newman:2006,richardson:mucha:porter:2009,ruan:zhang:2008} while progressively lowering the resolution limit may be able to find the appropriate refinement (although some problems can remain if the divisions are always binary~\cite{richardson:mucha:porter:2009}). Alternatively, the history of merges within agglomerative algorithms may provide a way to identify the intuitive modules that were merged to obtain the optimal partition~\cite{berry:etal:2009,clauset:etal:2004}. Multi-scale modularity-based methods~\cite{reichardt:bornholdt:2006,li:etal:2008,arenas:fernandez:gomez:2008} allow a researcher to specify a target resolution limit and identify modules on that scale, but it is typically not clear how to choose the ``correct'' target resolution {\em a priori}. Finally, Berry et al.~\cite{berry:etal:2009} recently showed that in some situations, the resolution limit can be circumvented with a clever edge-weighting scheme. These possibilities are encouraging, but most have yet to be fully characterized. 

More generally, this discussion of intuitive versus optimal partitions ignores two subtle problems in the general task of identifying modules from connectivity data alone. First, there is the choice of a random graph as the null model, which, as we showed above, plays a critical role in generating resolution limits. If we could instead choose a null model with more realistic assumptions about inter-module connectivity, unintuitive merges would become less likely. For instance, the null model assumes that an edge emerging from some module can, in principle, connect to any node in the network, but for real-world systems this assumption is rarely accurate (a point also recently made by Fortunato~\cite{fortunato:2010}).

A related issue is that the null model is unforgiving of sampling fluctuations, even those naturally generated by the null model itself. That is, the null model is a mean-field assumption, while actual networks --- even those drawn from the null model --- naturally deviate from these expected values. Such fluctuations are ultimately responsible for the non-zero maximum modularity scores observed in homogeneous random graphs~\cite{guimera:sales-pardo:amaral:2004}.
This issue is more severe in sparse networks, where the expected inter-module connectivity will tend to be less than one edge, while sampling alone will generate a non-trivial number of such connections. Modularity will see these connections as ``surprising'' and may mistake them for structure internal to a module. The Berry et al.~\cite{berry:etal:2009} edge reweighting approach can serve to dampen this effect by reducing the relative weight of inter-module edges so that they appear closer to what we expect under the null model. A more tolerant definition of modularity might only merge groups of nodes if their observed interconnectivity were statistically significant relative to the null model (an approach hinted at by Refs.~\cite{sales-pardo:etal:2007,ruan:zhang:2008,karrer:etal:2008,lancichinetti:etal:2009}).

Second, and more fundamentally, in order to distinguish an optimal partition from an intuitive partition, we must assume an external definition of an ``intuitive'' module. The fact that there exist instances where modularity maximization produces counter-intuitive results, i.e., results that clash with our external definition, simply highlights the difficulty of constructing a mathematical definition of a module that always agrees with our intuition. Indeed, it is unknown whether our intuition is even internally consistent.

Precisely the same difficulty lies at the heart of a decades-long and ongoing debate over how best to identify ``clusters'' in spatial data, which are conventionally understood to be non-trivial groups of points with high internal densities. For instance, Kleinberg recently proved that no mathematical definition of a spatial cluster can simultaneously satisfy three intuitive requirements~\cite{kleinberg:2002:b}, while Ackerman and Ben-David, taking a different approach, arrived at a contradictory conclusion~\cite{ackerman:ben-david:2008}. For identifying modules in complex networks, the debate has only just begun and it remains to be seen whether ``impossibility'' results from spatial clustering also apply to network clustering.

\section{Extreme Degeneracy Among High-Modularity Partitions} 
\label{section:degeneracies}
If we take the mathematically principled stance and accept modularity's definition of a good module, i.e., we do not assume any external notions, the modularity function still admits a subtle and problematic behavior for practical optimization techniques: even when it is not beneficial to merge two modules, i.e., when $\Delta Q_{ij}<0$, the penalty for doing so can be very small. Further, as the number of  modular structures increases, the number of ways to combine them in these suboptimal ways grows exponentially. Together, these properties lead to extreme degeneracies in the modularity function, which are problematic both for finding the maximum modularity partition and for interpreting the structure of any particular high-modularity partition. Thus, we have a highly counter-intuitive situation: as a network becomes more modular, the globally optimal partition becomes harder to find among the growing number of suboptimal, but competitive, alternatives.

\subsection{Modular networks}

To make this argument quantitative, consider a network composed of $k$ sparsely interconnected groups of nodes with roughly equal densities $d_{i} \approx 2m/k$. Even when $m$ is small enough that the intuitive partition coincides with the optimal partition (i.e., when there is no clash between our intuition and the definition of modularity), Eq.~\eqref{eq:merge-penalty} shows that the change in modularity for merging a pair of these groups is bounded below by $\Delta Q_{ij} = -2 k^{-2}$. For a moderate choice of $k=20$, this change is at most $\Delta Q_{ij} = -0.005$, which implies that these alternative partitions have modularities very close to $Q_\mathrm{max}$. As the number of groups $k$ increases, this penalty tends toward zero, and it becomes increasingly difficult for the modularity function to distinguish between the optimal partition and these suboptimal alternatives.

If there were only a few competitive alternatives, this degeneracy problem might be manageable. Unfortunately, their number grows combinatorially with the number of modular structures $k$. Its precise behavior depends on the inter-module connectivity, but is bounded below by $2^{k-1}$ and above by the $k$th Bell number.

The lower bound can be seen by considering the connected modular network with the fewest inter-module edges. This is given by the ``string'' network, which is a ring network with one inter-module edge removed. In this case, the number of suboptima is equal to the number of ways we can cut inter-module edges to divide the $k$ groups into connected components. Because there are $k-1$ such edges, each of which can be either cut or not cut, the number of partitions we can obtain this way is exactly $2^{k-1}$.

The upper bound comes from a network where each of the $k$ groups is connected to every other group, and the number of suboptima here is exactly equal to the number of ways to partition the $k$ groups into $k'$ groups of groups, for all choices of $k'$. This is given by the $k$th Bell number, which grows faster than exponentially.

The intermediate levels of degeneracy correspond to networks with varying degrees of inter-module connectivity: sparse modular networks will be closer to the lower bound, while dense modular networks will be closer to the upper bound.

At face value, this finding seems to contradict that of Massen and Doye~\cite{massen:doye:2006}, who argued that empirical networks with modular structure tend to exhibit a strong global peak around the optimal partition. This is a red herring. Recall that the total number of partitions of $n$ nodes grows like the $n$th Bell number, while the fraction of these that correspond to degenerate solutions is vanishingly small, since $k<n$. Thus, the modularity function is strongly peaked in a relative sense:  even a super-exponential number of degenerate, high-modularity solutions can still be a vanishingly small fraction of the enormous number of partitions in general.


\subsection{Hierarchical networks}

In addition to modular structure, many networks exhibit {\em hierarchical} structure, in which their nodes divide into groups that further subdivide into groups of groups, etc.\ over multiple scales~\cite{guimera:amaral:2005,sales-pardo:etal:2007,girvan:newman:2002,lagomarsino:etal:2007,clauset:etal:2008}, and where groups that are closer together in this hierarchy tend to be more densely interconnected. Here, we show that such networks exhibit at least as many degenerate solutions as simple modular networks, and that the modularity scores of alternative solutions can be even closer.

Suppose that the optimal partition of a hierarchical network contains two modules $i$ and $j$, each of which is composed of exactly two subgroups so that $i=\{a,b\}$ and $j=\{c,d\}$. Let us first split $i$ and $j$ into their constituent subgroups $\{a,b,c,d\}$ and then merge the opposite pairs of subgroups to obtain the suboptimal partition $i'=\{a,c\}$ and $j'=\{b,d\}$. From Eq.~\eqref{eq:merge-penalty}, the change in modularity $Q$ for this operation is exactly
\begin{align}
\Delta Q = & ~ (\Delta Q_{ac} + \Delta Q_{bd}) - (\Delta Q_{ab} + \Delta Q_{cd}) \nonumber \\
               = & ~ \frac{(e_{ac} + e_{bd}) - (e_{ab}+e_{cd})}{m} \nonumber \\
               & ~~- 2 \left( \frac{d_a - d_d}{2m} \right) \left( \frac{d_c - d_b}{2m} \right) \enspace .
\label{eq:split-merge-penalty}
\end{align}
%
Unlike Eq.~\eqref{eq:merge-penalty}, the size of the penalty now depends only on differences in connectivities, rather than on their absolute values, and will thus tend to be much smaller than the penalty 
discussed in the previous section.

If the network's hierarchical structure is relatively balanced (i.e., submodules at the same level in the hierarchy have similar degree $d$) then Eq.~\eqref{eq:split-merge-penalty} will be dominated by its first term, whose size depends only on the differences in the pairwise connectivities of the submodules.  This is very small both when the groups $i$ and $j$ are close to each other in the hierarchy, e.g., are siblings or cousins, and thus have similar inter- and intra-module connectivities, and when $i$ and $j$ are relatively low in the hierarchy, and thus have few connections to begin with.

Furthermore, each level of a hierarchy presents its own set of modular structures that can be merged, either within the same level or between different levels of the hierarchy. The number of ways these structures can be combined depends on their particular hierarchical organization and the number of connections between distantly related groups. Generally, however, it follows the same bounds we showed above for a non-hierarchical network, i.e., at least $2^{k-1}$ and no more than the $k$th Bell number, when there are $k$ modular structures at the lowest level of the hierarchy.

We also note that hierarchical problems are not, in fact, limited to hierarchical networks.  The resolution limit phenomenon, which tends to produce agglomerations with modular substructure, creates effective hierarchical structure even in a non-hierarchical network and thus can induce hierarchy-style degeneracies in the modularity function.

In both cases considered above, the existence of extreme degeneracies in the modularity function does not depend on the detailed structure of the particular network or on any external notion of a ``true'' module. Instead, these solutions exist whenever a network is composed of many groups of nodes with relatively few inter-group connections. In a sense, these groups constitute the ``building blocks'' used to construct the high-modularity partitions.

Fundamentally, these degeneracies arise because the modularity function does not strongly penalize partitions that combine such groups and the degeneracies are legion because there are at least an exponential number of such combinations. As a consequence, the modularity function is not strongly peaked around the optimal partition---in physics parlance, the modularity function is {\em glassy}---in precisely the case that we would like modularity maximization to perform best: on modular networks. 

We note that similar degeneracies are likely to occur in other kinds of module-identification quality functions, including some likelihood-based functions~\cite{bickel:chen:2009} and they persist under directed, weighted, bi-partite and multi-scale generalizations of modularity. For the $\gamma$-generalization of $Q$ introduced by Reichardt and Bornholdt~\cite{reichardt:bornholdt:2006}, choosing $\gamma<1$ increases the severity of the degeneracy problem, by reducing the penalty for merging modules, while choosing $\gamma>1$ reduces it somewhat, by increasing the penalty. For any fixed $\gamma$, however, there exist many networks that will exhibit severe degeneracies, and, moreover, it remains unclear how to identify the ``correct'' value of $\gamma$ without resorting again to an external definition of a ``true'' module. Similar issues apply to other parametric generalizations of modularity~\cite{li:etal:2008,arenas:fernandez:gomez:2008}. For most weighted networks, the degeneracy problem will tend to be stronger because weights effectively reduce the penalty for merging some modules. Also, many weighted networks are dense and, as we showed above, these exhibit many more degenerate solutions than sparse networks (even if they may not necessarily exhibit a resolution limit; see Section~\ref{section:resolution:limit}).

%
%

Of course, an optimal partition always exists, even if it may be almost impossible to find in practice. But the scientific value of the optimum does seem somewhat diminished by the enormous number of structurally diverse alternatives that are only slightly ``worse'' from the perspective of their modularity scores. That is, without external information, it becomes unclear which particular partition, within the enormous number of roughly equally good ones, is more scientifically meaningful~\cite{agarwal:kempe:2008}. And, requiring such external information defeats the purpose of identifying modules automatically from connectivity data alone.

\section{The Limiting Behavior of $Q_{\max}$ for Modular Networks}
\label{section:limiting:behavior}

In addition to the location of the peak of the modularity function and its surrounding structure, another important question is the expected height of the peak. In this section we show that, in the asymptotic limit of an increasingly modular network --- i.e., as we add more modules to the network --- the height of the modularity function converges to $Q_{\max}=1$. This analysis fills an important gap in our understanding of the modularity function's behavior in practical contexts, as previously only unrealistic cases such as lattices, trees and Erd\"{o}s-R\'enyi random graphs~\cite{guimera:sales-pardo:amaral:2004, reichardt:bornholdt:2006b} have been considered. Notably, these results serve to explain why large values of $Q_{\max}$ are often found for very large real-world networks~\cite{blondel:etal:2008}.

Consider a sparse network with $n$ nodes, $m = O(n)$ edges and an optimal partition that contains $k$ modules. Because modularity is a summation of contributions from individual terms, we may rewrite Eq.~\eqref{eq:modularity} for the optimal partition as
\begin{align}
Q_{\max} & = \sum_{i=1}^{k} \left[ \frac{\langle e \rangle}{m} - \Big \langle \left( \frac{d_i}{2m} \right)^2 \Big \rangle \right] \nonumber \\
    & = \sum_{i=1}^{k} \left[ \frac{\langle e \rangle}{m} - \left( \frac{\langle d \rangle}{2m} \right)^2 - \mathrm{Var}\!\left(\frac{d_i}{2m}\right) \right] \enspace , \label{eq:qmax}
\end{align}
where $\langle e \rangle = \frac{1}{k} \sum_i e_i$ is the average number of edges within an optimal module, $\mathrm{Var}(.)$ is the variance function and $\langle d \rangle = \frac{1}{k} \sum_i d_i$ is the average degree of an optimal module.

Now, imagine a process by which we connect new modular subgraphs of some characteristic size $\langle e \rangle = O(1)$ to the network, i.e., we assume that the average size of a module does not increase as the network grows (but see below), and consider the asymptotic dependence of $Q_{\max}$ in the limit of this infinitely modular network.

It will be convenient to rewrite the average degree of a module as
\begin{align}
\langle d \rangle & = 2 \langle e \rangle + \langle e^{\rm out} \rangle \enspace , \label{eq:mean:d}
\end{align}
where $\langle e^{\mathrm{out}} \rangle = \frac{1}{k} \sum_i e^{\mathrm{out}}_i$ denotes the average number of outgoing edges in each module. Because modules do not grow with the size of the network, the number of modules $k$ is $O(n)$, and hence the average out-degree $\langle e^{\mathrm{out}} \rangle$ must be $O(1)$ to ensure that the network remains sparse.  This implies that the average degree $\langle d \rangle$ is also $O(1)$. Finally, because ${\rm Var}(d_i/2m) \to 0$ in the limit, we may ignore the last term in Eq.~\eqref{eq:qmax}.

Combining the expression for $\langle d \rangle$ [Eq.~\eqref{eq:mean:d}] with the expression for the maximum modularity [Eq.~\eqref{eq:qmax}], we have
\begin{align}
Q_{\max}   & = \sum_{i=1}^{k} \left[ \frac{\langle e \rangle}{m} - \left( \frac{2 \langle e \rangle + \langle e^{\rm out}\rangle}{2m} \right)^2 \right] \enspace . \label{eq:qmax:2}
\end{align}
Rewriting the number of edges in the network $m$ as a function of the connectivity of the optimal modules
\begin{align}
m & = \frac{1}{2} k \langle d \rangle = \frac{k}{2}(2\langle e \rangle + \langle e^{\mathrm{out}} \rangle) \enspace , \nonumber
\end{align}
and carrying out the summation in Eq.~\eqref{eq:qmax:2}, we see that 
\begin{align}
Q_{\max} & =  \frac{\langle e \rangle}{\langle e \rangle + \langle e^{\mathrm{out}} \rangle /2} - \frac{1}{k} \nonumber \\
    & = \frac{1}{1 + \frac{ \langle e^{\mathrm{out}} \rangle }{ 2 \langle e \rangle }} - \frac{1}{k} \enspace . \label{eq:qmax:penultimate}
\end{align}
Thus, in the limit of $k\to\infty$, $Q_{\max}$ approaches some constant less than $1$, which depends only on the relative proportion of internal to external edges in each module. However, this analysis is incomplete in a crucial way: it ignores the impact of the resolution limit described in Section~\ref{section:resolution:limit}, which can cause the average size of a module in the optimal partition to grow with the size of the network~\cite{fortunato:barthelemy:2007,berry:etal:2009,kumpula:etal:2007,branting:2008}.

When the resolution limit causes two groups of nodes to be merged, the links joining them become internal, which in the limit causes the average out-degree of a module in the optimal partition to be asymptotically dominated by its average internal density, i.e., $\langle e^{\mathrm{out}} \rangle=o(\langle e \rangle)$. This, in turn, implies that $\langle e^{\rm out} \rangle / \langle e \rangle \to 0$ as $k\to\infty$. Accounting for this resolution-limit induced agglomeration, we now see that the first term in Eq.~\eqref{eq:qmax:penultimate} approaches $1$ while the second term approaches $0$, implying that $Q_{\max} \to 1$ as $k \to \infty$.
[Recall, however, that this last step does not hold for all weighted networks: consider a limiting process in which each module connects to $O(k)$ other modules with total weight $o(k)$, e.g., the collection of $k$ cliques connected to each other by edges of weight $2/(k-1)$ from Section~\ref{section:resolution:limit}. The resolution is not present in such a network and hence $Q_\mathrm{max}$ merely approaches the value given in Eq.~\eqref{eq:qmax:penultimate}.]

Thus, just as the severity of the degeneracy problem depends strongly on the number of modular structures in the network, so too does the height of the modularity function. Further, the number of these structures $k$ is limited mainly by the size of the network, since there cannot be more modular structures than nodes in the network. In practical contexts, variations in $n$ are very likely to induce variations in $Q_{\max}$ and increasing $n$ (or $k$) will generally tend to increase $Q_{\max}$. If the intention is to compare modularity scores across networks, these effects must be accounted for in order to ensure a fair comparison.

Of course, the precise dependence of $Q_{\max}$ on $n$ and $k$ depends on the particular network topology and how it changes as $n$ or $k$ increases. For instance, in Appendix~\ref{appendix:qmax:ring}, we derive the exact dependence for the ring network and calculate precisely how many of its degenerate solutions lie within 10\% of $Q_{\max}$. Because of this dependence, an estimate of $Q_{\max}$ for any empirical network should not typically be interpreted without a null expectation based on networks with a similar number of modules. For instance, detailed values of $Q$ should probably not be compared across different networks, as in a regression of modularity $Q$ versus network size $n$~\cite{kreimer:etal:2008}.

Finally, we point out that this dependence of $Q_{\max}$ on $n$ and $k$ makes intuitive sense given that the null model against which the internal edge fractions of the modules are scored [the second term in Eq.~\eqref{eq:modularity}] is a random graph with the same degree sequence (see Section~\ref{section:resolution:limit}). That is, as the number of modules increases, it is increasingly unlikely under the null model that any edges fall within a particular module given the huge number of possible connections to other modules. In this sense, it is not at all surprising that extremely high modularity values have been found for extremely large real-world networks. For instance, Blondel et al.~\cite{blondel:etal:2008} estimated $Q_{\max}\geq0.984$ for one Web graph with $118$ million nodes and $Q_{\max}\geq0.979$ for a different Web graph with $39$ million nodes. Such high values may not indicate that they are particularly modular, but instead that they are simply very different from a random graph with the same degree sequence.

\section{Mapping the Modularity Landscape}
\label{sec:mapping:landscape}
To get a more intuitive handle on the precise structure of the modularity function, we now describe a numerical technique for reconstructing a locally accurate, low-dimensional visualization of it. We then apply this technique to instances of synthetic modular or hierarchical networks. In the next section, we apply it to three real-world networks and show that the high-modularity partitions of empirical networks can disagree strongly on many, but not all, partition properties.

\begin{figure}[t]
\begin{center}
\includegraphics[scale=0.25]{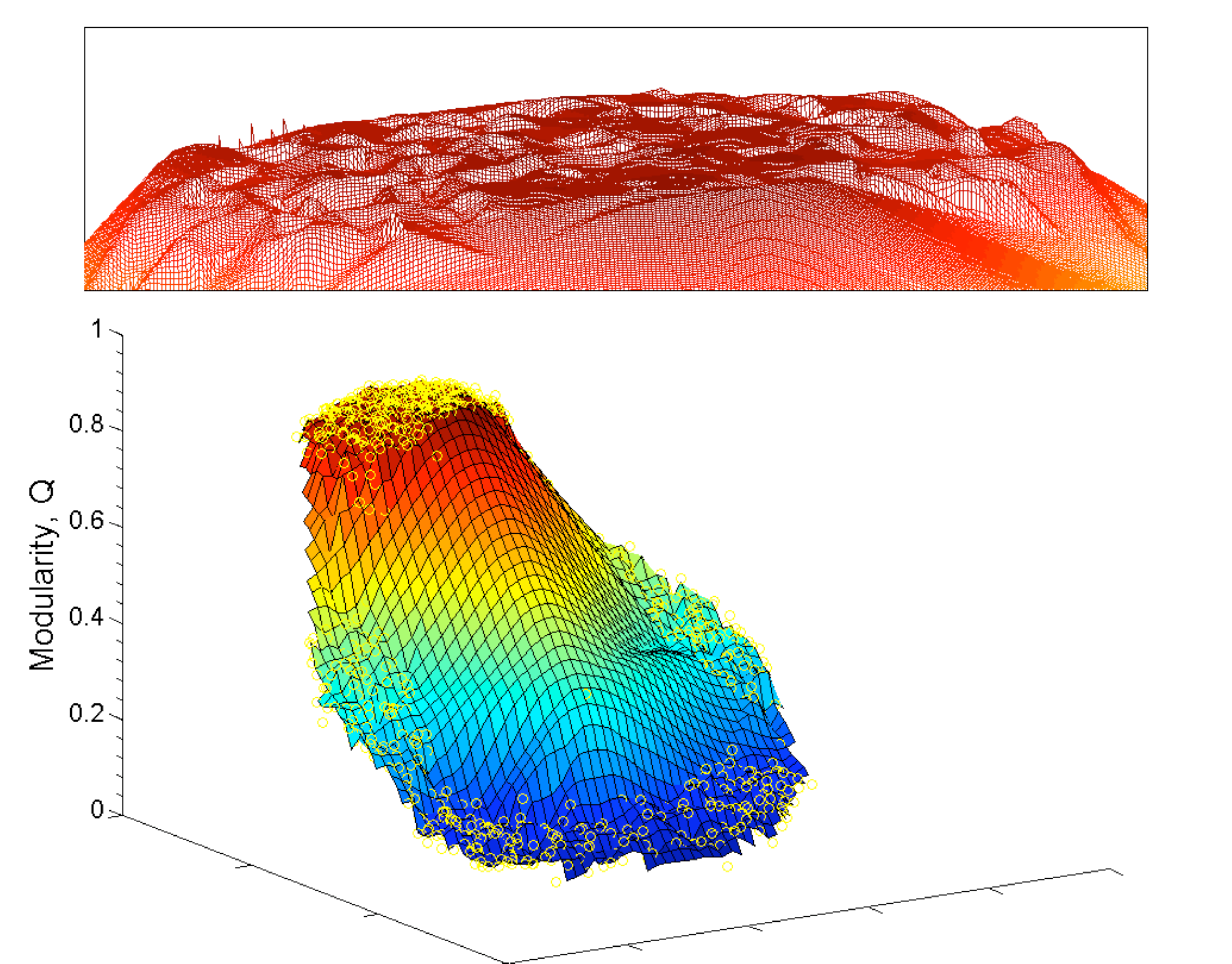}
\caption{(color online) The modularity function of a ring network ($k=24$ and $c=5$), reconstructed from 997 sampled partitions (circles), showing a prominent high-modularity plateau. The vertical axis gives the modularity $Q$; the $x$- and $y$-axes are the embedding dimensions. (These dimensions are complicated functions of the original partition space and thus their precise scale is not relevant; see Appendix~\ref{appendix:embedding}). Note that the structure within the plateau (inset) is highly irregular, illustrating the severe degeneracies of the modularity function. }
\label{fig:ring:partitions}
\end{center}
\end{figure}

\subsection{The reconstruction technique}

Our focus here is on the modularity function's structure in the vicinity of the degenerate high-modularity partitions identified in Section~\ref{section:degeneracies}. To do this, we sample partitions using a simulated annealing (SA) algorithm. Each SA sample was started from a random initial partition and stopped either at a randomly chosen step (75\% of runs) or at a local optimum (25\% of runs). This mixture of stopping points ensures that our sample contains both a large number of local optima as well as a sampling of sub-optimal partitions in their vicinity. Complete details of the sampling approach are given in Appendix~\ref{appendix:sa}.

To reconstruct and visualize the structure implied by these sampled partitions, we embed them as points in a 2-dimensional Euclidean space such that we largely preserve their pairwise distances. The distance between partitions is measured by one popular distance metric for partitions, called the variation of information (${\rm VI}$)~\cite{meila:2005} and defined as follows. Given partitions $C$ and $C'$, the variation of information between them is defined as
\begin{align}
{\rm VI}(C,C') & = H(C,C') - I(C;C') \enspace ,
\label{eq:vi-first}
\end{align}
where $H(.;.)$ is the joint entropy [see Eq.~\eqref{eq:joint:entropy}] and $I(.;.)$ is the mutual information [see Eq.~\eqref{eq:mutual:information}] between the two partitions. Additional details 
are given in Appendix~\ref{appendix:distances}. We note that using other measures of partition distance, such as one based on the Jaccard coefficient, yields similar results (see Fig.~\ref{fig:metabolic:jaccard} in Appendix~\ref{appendix:distances}).

\begin{figure}[t]
\begin{center}
\includegraphics[scale=0.25]{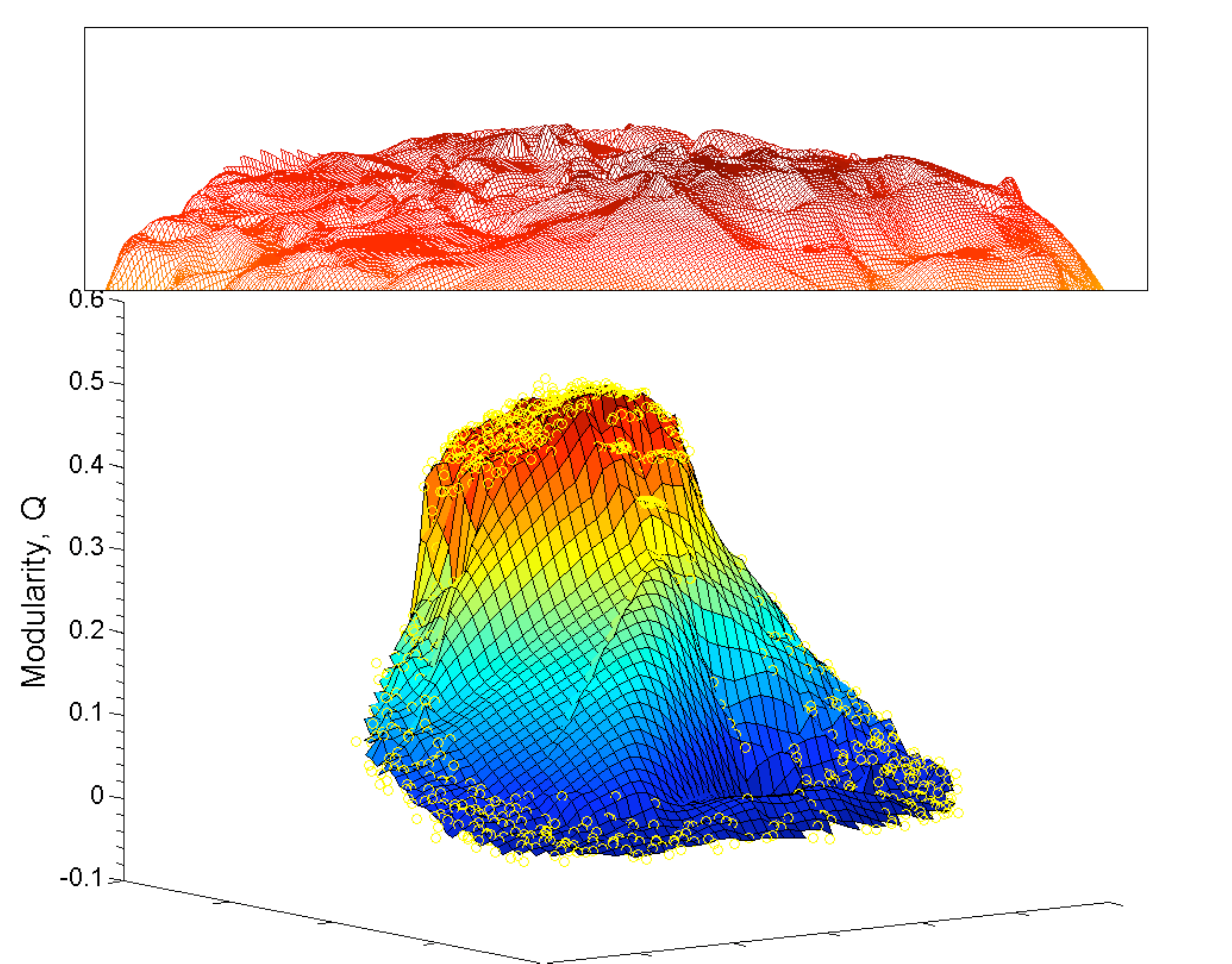}
\caption{(color online) The modularity function of a hierarchical random graph model~\cite{clauset:etal:2008}, with $n=256$ nodes arranged in a balanced hierarchy with assortative modules (see Appendix~\ref{appendix:hrg}), reconstructed from 1199 sampled partitions (circles), and its rugged high-modularity region (inset). }
\label{fig:hrg:partitions}
\end{center}
\end{figure}

The embedding portion of our reconstruction technique seeks an assignment of partitions $\{C_{i}\}$ to coordinates $\{(x_{i},y_{i})\}$ such that the pairwise distances between partitions are largely the same as the pairwise distances between embedded points. Only the relative positions of points in the embedded space are significant; their precise locations are meaningless. Then, by assigning each embedded point a value in a third dimension equal to the modularity score $Q_{i}$ of the corresponding partition, we can directly visualize the structure of the sampled modularity function.

Because we are interested in the function's degeneracies, we prefer an assignment that errs on the side of being more smooth, i.e., less rugged, in the projected space than in the original partition space. Methods like principal component analysis use a linear function to measure the quality of the embedding, which can cause some local structure to be lost or distorted when projecting from non-Euclidean spaces like the partition space. Instead, we use a technique called curvilinear component analysis (CCA)~\cite{lee:verleysen:2007}, which preserves local distances at the expense of some distortion at larger distances. Thus, our reconstructed modularity landscapes are appropriately conservative, sometimes reducing the apparent ruggedness of the reconstructed landscape, but never creating ruggedness where it does not exist in the first place.

Additional details of the CCA technique are given in Appendix~\ref{appendix:embedding}. For completeness, we note that several other approaches to mapping specific features of the modularity landscape are described in Refs.~\cite{sales-pardo:etal:2007,sawardecker:sales-pardo:amaral:2009,holmstrom:bock:brannlund:2009}.

\subsection{Reconstructed modularity functions for modular and hierarchical networks}

Using a ring network with $k=24$ and $c=5$ (Fig.~\ref{fig:ring}), Fig.~\ref{fig:ring:partitions} shows the modularity function reconstructed from nearly 1000 sampled partitions.  Examining these in detail, we see that every low-modularity partition divides many cliques across different groups, which leads to low values of $Q$. In contrast, the high-modularity partitions are composed of various groupings of the cliques, as predicted in Section~\ref{section:degeneracies}. In the embedded modularity function, these high-modularity partitions tend to cluster together, forming a distinct ``plateau'' region. Within this region, the function shows complicated degeneracies and no clear maximum (Fig.~\ref{fig:ring:partitions}, inset).

Now turning to the case of a hierarchically structured network, we use a simplified version of the recently introduced hierarchical random graph (HRG) model~\cite{clauset:etal:2008}, in which we organize $n=256$ nodes into nested modules using a balanced binary tree structure---so that submodules at the same level in the hierarchy have similar sizes---and an assortative connectivity function--- so that submodules become more internally dense as we descend the hierarchy from large to small groups. Appendix~\ref{appendix:hrg} gives the precise details of the HRG model we use and analytically derives its optimal partition.

\begin{figure}[t!]
\begin{center}
\includegraphics[scale=0.25]{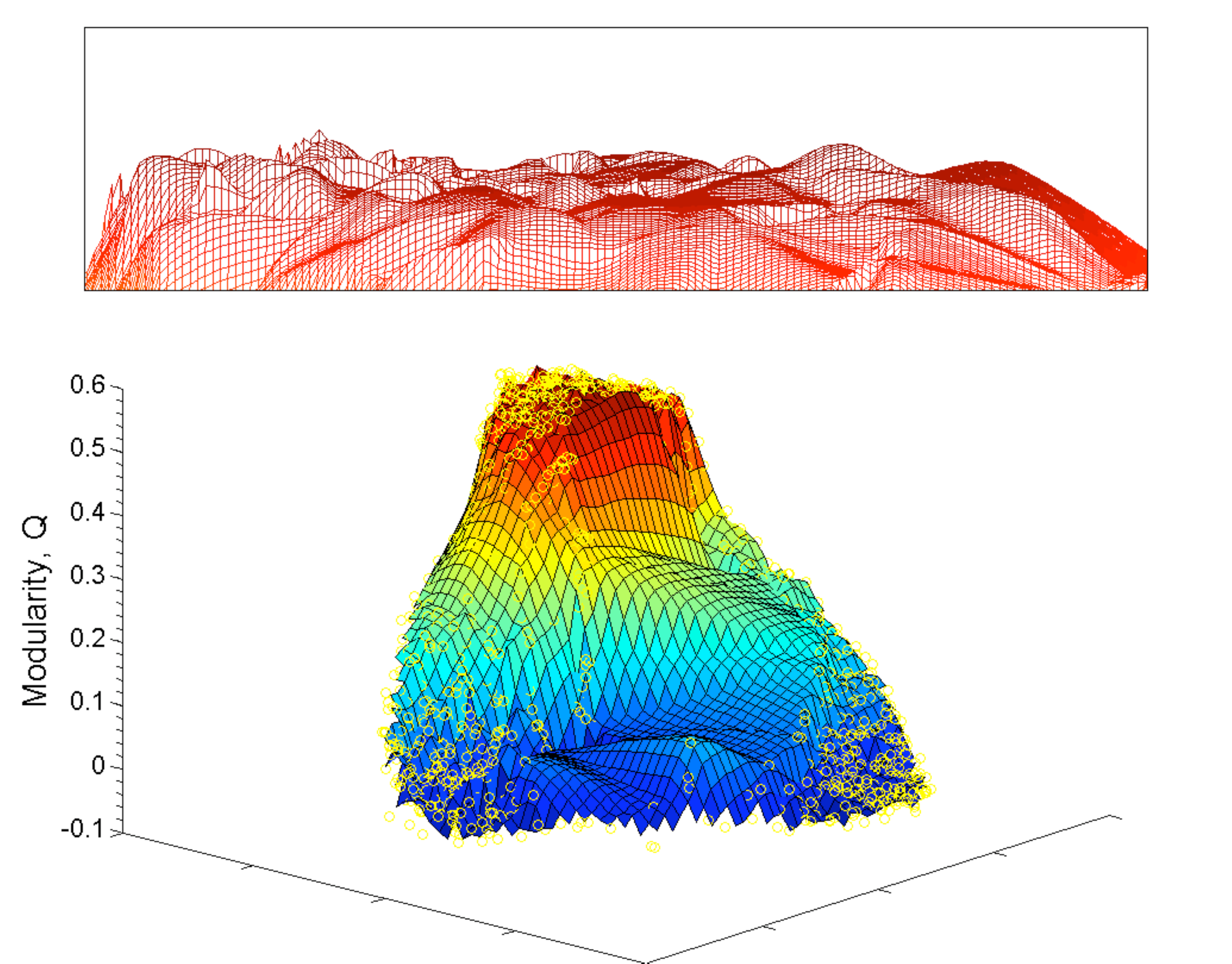} 
\caption{(color online) The modularity function for the metabolic network of the spirochaete {\em Treponema pallidum} with $n=482$ nodes (the largest component) and 1199 sampled partitions, showing qualitatively the same structure as we observed for hierarchical networks. The inset shows the rugged high-modularity region.}
\label{fig:metabolic:landscapes}
\end{center}
\end{figure}

From this model, we drew 100 network instances and combined sampling results from this ensemble to smooth out deviations caused by fluctuations in the random graph structure~\cite{massen:doye:2006,guimera:sales-pardo:amaral:2004}. As a consequence, the reconstructed modularity function is smoother than it would be for any particular instance and shows only the structure induced by the hierarchical organization of the network. Figure~\ref{fig:hrg:partitions} shows the reconstructed modularity function for nearly 1200 sampled partitions. Again examining these partitions in detail, we find that nearly all of the high-modularity partitions in the ``plateau'' region mix submodules from different levels of the hierarchy and often fail to resolve distinct branches, as predicted in Section~\ref{section:degeneracies}. Like the ring network (Fig.~\ref{fig:ring:partitions}), the high-modularity region in this case is extremely rugged, with many peaks and valleys and no clear global optimum (Fig.~\ref{fig:hrg:partitions}, inset).

As mentioned above, the CCA embedding technique only guarantees a lower bound on the ruggedness of the reconstructed modularity function. Thus, what appear to be local minima in the embedding are actually quite likely to be local maxima themselves and the true ruggedness is almost surely more extreme than it appears in these visualizations.

\section{Structural Diversity Among High-Modularity Partitions}
\label{section:realworld:networks}

Although our analytic arguments are entirely general, our numerical results have focused on specific synthetic networks derived from models of modular and hierarchical networks. These models may not be representative of the networks found in the real world, since they lack certain properties commonly observed in real-world networks (to name a few simple properties: unequal module sizes and heavy-tailed degree distributions~\cite{clauset:etal:2009}). In this section, we apply our reconstruction technique to several real-world examples of complex metabolic networks and consider the degree to which different high-modularity partitions agree on the large-scale modular structure.

\begin{figure}[t]
\begin{center}
\includegraphics[scale=0.55]{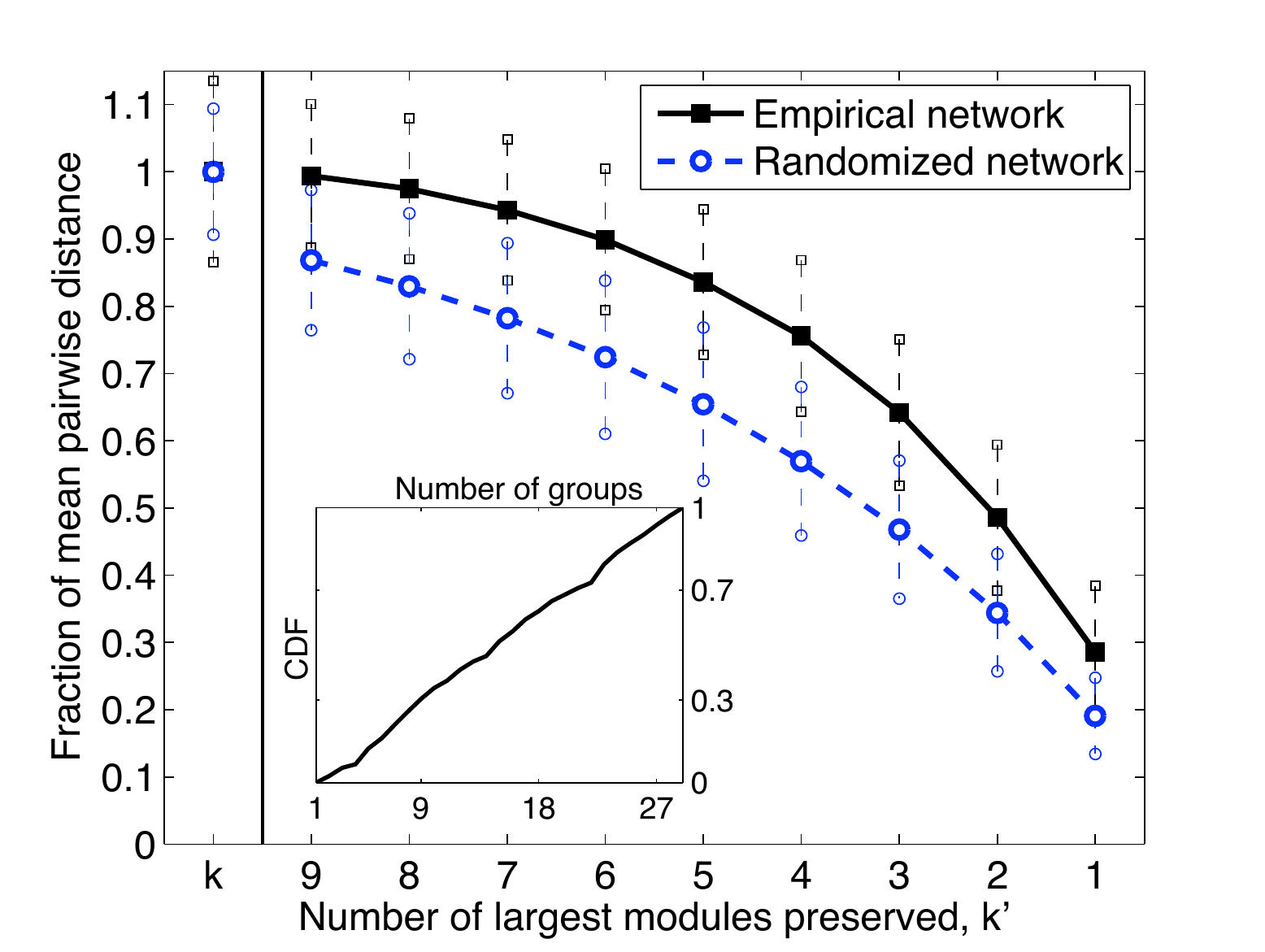} 
\caption{(color online) The fraction remaining of the mean pairwise distance between sampled high-modularity partitions when all but the $k'$ largest groups in each partition are merged into a single group, for the {\em T. pallidum} network and for a random graph with the same degree sequence. (Error bars indicate one standard deviation; inset shows the distribution of the number of groups in a partition.) In both cases, the fraction converges slowly on 0 as more modules are merged, indicating that the majority of the structural diversity captured by these partitions is driven by significant differences in the composition of the largest few identified groups. This behavior is especially true for the empirical network.
}
\label{fig:metabolic:coarse}
\end{center}
\end{figure}

Metabolic networks are an interesting test case for this analysis because the answers to many questions in systems biology depend on our ability to accurately characterize their modular and hierarchical structure~\cite{hartwell:etal:1999,barabasi:oltvai:2004,papin:reed:palsson:2004} and modularity maximization has been used extensively in their analysis. We emphasize, however, that our results likely also hold for other types of networks, such as social or technological networks, since the ruggedness of the modularity landscape depends only on the presence of modular or hierarchical structure.

Figure~\ref{fig:metabolic:landscapes} shows the reconstructed modularity function for the largest connected component in the  metabolic network of the spirochaete {\em Treponema pallidum} ($n=482$, $m=1223$) and Fig.~\ref{fig:metabolics:more} in Appendix~\ref{appendix:metabolic:networks} shows the functions for the mycoplasmatales {\em Mycoplasma pneumoniae} and {\em Ureaplasma parvum}~\cite{huss:holme:2007}. All three modularity functions are similar to those of the modular and hierarchical model networks shown in Figs.~\ref{fig:ring:partitions} and~\ref{fig:hrg:partitions}, exhibiting a broad and rugged region of high-modularity partitions with no clear global maximum.

\subsection{Large-scale similarity}

From a pragmatic perspective, the multiplicity of high-modularity partitions is more troublesome if they disagree on the large-scale modular structure of the network. In contrast, if high-modularity partitions disagree mainly on the composition of the smallest few modules, but agree on the composition of the larger ones, modularity maximization can provide useful information about a network's large-scale modular structure in spite of the degeneracies.

Using our sampled partitions, a direct and straightforward test of this possibility is the following. For each locally optimal partition, we set aside the $k'$ largest identified modules and then merge the remaining smaller modules into a single group. If most of the differences between local optima are in the composition of the smaller modules, the mean pairwise distance between the reduced partitions will vanish as we merge more of these small modules into a single group. However, if a significant fraction of the original mean pairwise distance remains even when almost all of the smaller modules have been merged, i.e., when $k'$ is small, then we have significant evidence that the high-modularity partitions fundamentally disagree on the networks' large-scale modular structure.

Figure~\ref{fig:metabolic:coarse} shows the results of this test using the sampled partitions of the {\em T. pallidum} metabolic network, for $9 \geq k' \geq 1$ (with similar results for the other metabolic networks; see Fig.~\ref{fig:metabolics:sensitivity} in Appendix~\ref{appendix:metabolic:networks}). For comparison, we also show results for a random graph with the same degree sequence, which has no real modular structure. Notably, the mean pairwise distance among the original empirical partitions decreases very little (0.05\%) when we retain only the $k'=9$ largest groups; in contrast, the random graph exhibits a much larger change (13\%).

Counter-intuitively, this implies that the high-modularity partitions of the random graph exhibit greater agreement on the composition of the largest few modules, i.e., less structural diversity, than do the high-modularity partitions of the empirical network. Additionally, the mean pairwise distance for the {\em T. pallidum} partitions only falls below 50\% of its original value when we merge all but the $k'=2$ largest groups. That is, almost half the variation of information between high-modularity partitions is explained by differences in the composition of their two largest modules, with the remainder being caused by disagreements on the composition of all other modules.

Thus, partitions that are ``close'' in terms of their modularity scores can be very far apart in terms of their partition structures and most of the differences come from disagreements on the composition of the largest identified modules. This suggests that the degeneracies in the modularity function really do pose a problem for interpreting the structure of any particular partition and that a high modularity score provides very little information about the underlying modular structure.

\subsection{Structural summary statistics}

For some research questions, however, the precise composition of the modules is not as important as the value of some statistical summary of the partition's structure. Thus, an important question is whether high-modularity partitions tend to agree on the values of simple summary statistics, even if they disagree on the precise partition structure. Naturally, the particular statistical quantity will depend on the research question being asked and the safest approach is to directly test whether the quantity measured on one high-modularity partition is representative of its distribution over many high-modularity partitions. Here, we briefly study two such summary statistics: the mean module density and the distribution of module sizes. We note, however, that tests of reliability like these may not generalize to larger networks, as the number of degenerate solutions, and thus their potential structural diversity, grows rapidly with the size of the network tested (see Section~\ref{section:degeneracies}).

\begin{figure}
\begin{center}
\includegraphics[scale=0.55]{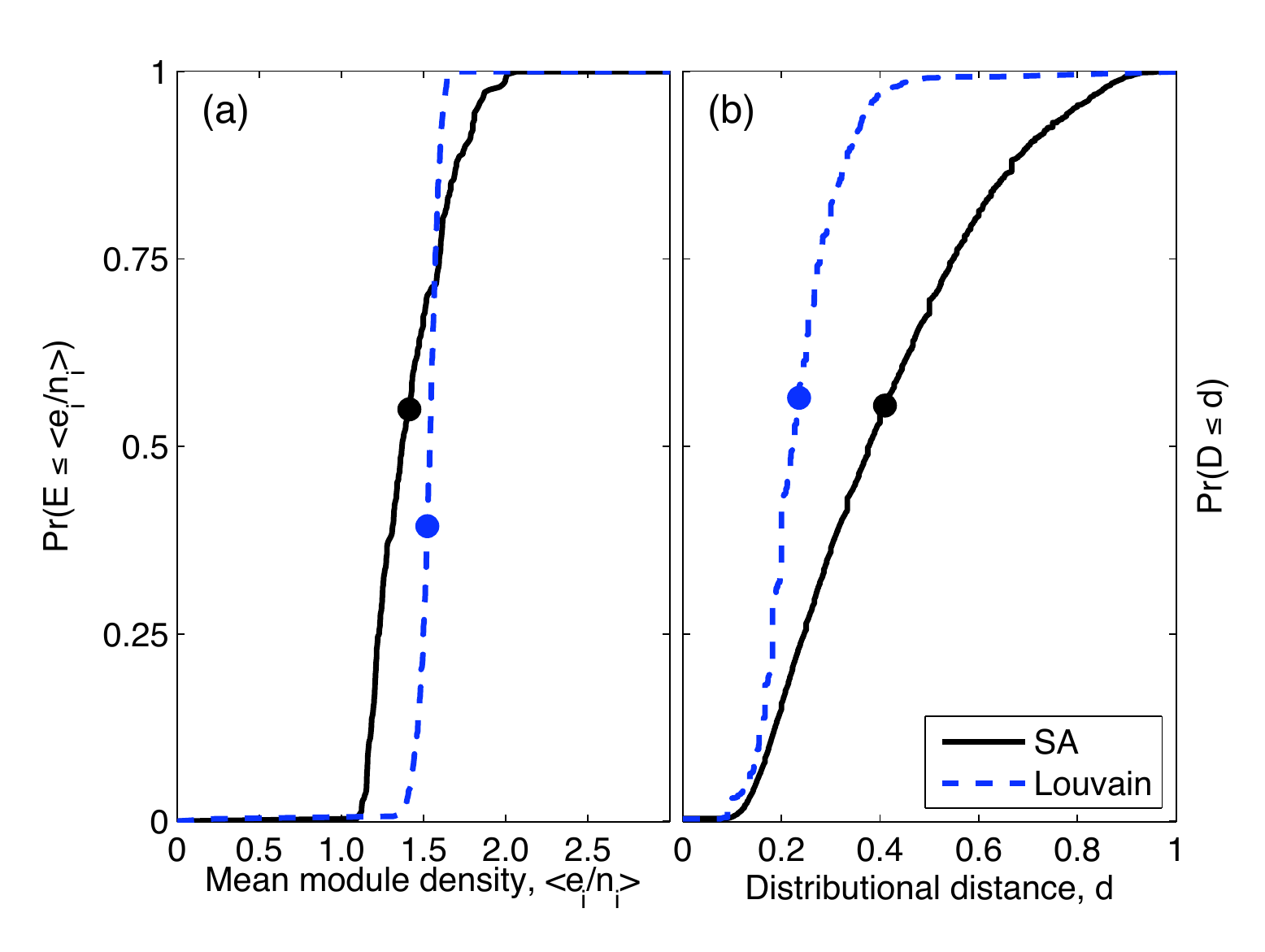} 
\caption{(color online) The cumulative distribution functions for the (a) mean module density $\langle e_{i}/n_{i} \rangle$ and (b) Kolmogorov-Smirnov distance $d$ between module size distributions $p(n_{i})$, among sampled high-modularity partitions of the {\em T. pallidum} network. (A dot indicates the distributional mean.) In both cases, we show the distributions for partitions derived using simulated annealing and using the Louvain method. In both cases, the SA partitions exhibit a much less tightly peaked distribution than those derived using the Louvain method, indicating that high-modularity partitions exhibit non-trivial structural diversity, even under these summary statistics. }
\label{fig:metabolic:summaries}
\end{center}
\end{figure}

Using the same high-modularity partitions of the {\em T. pallidum} metabolic network, along with a second set of high-modularity partitions derived using the Louvain method~\cite{blondel:etal:2008}, we compute the average module density $\langle e_{i}/n_{i}\rangle $ and the distribution of module sizes $p(n_{i})$ for each partition. The former statistic can be immediately compared between partitions; to compare the latter, we compute pairwise Kolmogorov-Smirnov (KS)~\cite{press:etal:1992} distances between the distributions. If the statistic's distribution is tightly concentrated, then any particular high-modularity partition can be assumed structurally representative, under that summary statistic, of the other high-modularity partitions.

Figure~\ref{fig:metabolic:summaries} shows the resulting distributions for our two simple measures. In both cases, the distributions for the Louvain partitions are indeed relatively tightly concentrated, illustrated by a large increase of the CDF over a small range in the $x$ variable. This suggests that the Louvain method tends to find partitions with relatively similar module densities and module sizes. In contrast, however, the SA partitions exhibit much more variance under both measures. This indicates that the SA method more accurately samples the full structural diversity of the high-modularity partitions than does the Louvain method. (To be fair, the Louvain method was not designed to find representative high-modularity partitions, but rather to be very fast at finding some high-modularity partition.)
%
%

On this network, both methods tend to produce partitions with similar mean module densities: the estimated means are $\langle e_{i}/n_{i}\rangle=1.421 \pm 0.013$ [mean$\pm$std.\ err.] for SA versus $1.520 \pm 0.005$ for Louvain. The Louvain method, however, underestimates this statistic's standard deviation by about a factor of 2 relative to the SA method ($\sigma = 0.223$ for SA versus $\sigma = 0.095$ for Louvain).

Thus, these results support our conclusion above: high-modularity partitions can exhibit non-trivial structural diversity, even under simple structural measures like the mean module density and the distribution of module sizes. Of these two, the mean module density is more reliably representative, although even it exhibits non-trivial variance. In contrast, the distribution of module sizes exhibits a great deal of variation.  Thus, we generally recommend a cautious approach when interpreting the structure of one or a few high-modularity partitions, as their structural characteristics may not be representative of alternative high-modularity solutions.

\section{Discussion}
To summarize, the modularity function $Q$ poses three distinct problems in scientific applications: 
\begin{enumerate}
\item The optimal partition may not coincide with the most intuitive partition (the resolution limit problem~\cite{fortunato:barthelemy:2007,berry:etal:2009,kumpula:etal:2007,branting:2008}), an effect driven primarily by the consequences of assuming that inter-module connectivity follows a random graph model (see Section~\ref{section:resolution:limit}).
\item There are typically an exponential number of structurally diverse alternative partitions with modularities very close to the optimum (the degeneracy problem). This problem is most severe when applied to networks with modular structure; it occurs for weighted, directed, bi-partite and multi-scale generalizations of modularity; and it likely exists in many of the less popular partition score functions for module identification (see Sections~\ref{section:degeneracies},~\ref{sec:mapping:landscape} and~\ref{section:realworld:networks}).
\item The maximum modularity score $Q_{\max}$ depends on the size of the network $n$ and on number of modules $k$ it contains (see Section~\ref{section:limiting:behavior}).
\end{enumerate}
To be practically useful, we believe that future methodological work on module identification in complex networks must, in particular, include some effort to address the existence of degenerate solutions and the problems they pose for interpreting the results of the procedure.

The discovery of extreme degeneracies in the modularity function also provides an answer to a nagging question in the literature: given that maximizing modularity is NP-hard in general~\cite{brandes:etal:2008}, why do so many different heuristics perform so well at maximizing it in practice? And further, why do different methods often return different partitions for the same network? The answer is that the exponential number of high-modularity solutions makes it easy to find some kind of high-scoring partition, but, simultaneously, their enormous number obscures the true location of the optimal partition.

In this light, it is unsurprising that different heuristics often yield different solutions for the same input network, particularly for very large networks. Different heuristics will naturally sample or target distinct subsets of the high-modularity partitions due to their different approaches to searching the partition space (for instance, see Fig.~\ref{fig:algorithm-compare}, in Appendix~\ref{appendix:sa}). This implies that the results of deterministic algorithms, such as greedy optimization~\cite{newman:2004:b,clauset:etal:2004,blondel:etal:2008} or spectral partitioning~\cite{newman:2006,richardson:mucha:porter:2009}, which return a unique partition for a given network, should be treated with particular caution, since this behavior tends to obscure the magnitude of the degeneracy problem and the wide range alternative solutions.

The structural diversity of high-modularity partitions (Figs.~\ref{fig:metabolic:coarse} and~\ref{fig:metabolic:summaries}) suggests that a cautious stance is typically appropriate when applying modularity maximization to empirical data.  Unless a particular optimization or sampling approach can be shown to reliably find representative high-modularity partitions, the precise structure of any high-modularity partition or statistical measures of its structure should not be completely trusted. 

Finally, even the estimated modularity score $Q_{\max}$, which may be ``significant'' relative to a simple random graph~\cite{guimera:sales-pardo:amaral:2004}, should be treated with an ounce of caution as it is almost always a lower bound on the maximum modularity (but see Ref.\ \cite{agarwal:kempe:2008}) and its accuracy necessarily depends on the particular algorithm and network under consideration. As a result, an estimate of $Q_{\max}$ should not be mistaken for a network property that can be fairly compared across networks: as we showed in Section~\ref{section:limiting:behavior}, $Q_{\max}$ depends on the number of module-like structures in the network and on their interconnectivity, both of which are limited by the network's size. Thus, variation in size can induce variation in the maximum modularity value and a fair comparison between different networks must control for this correlated behavior.

Although the degeneracy problem presents serious issues for the use of modularity maximization in scientific contexts, certain kinds of sophisticated approaches may be able to circumvent or mitigate some of its consequences. For example, Sales-Pardo et al.~\cite{sales-pardo:etal:2007} recently proposed combining information from many distinct high-modularity partitions to identify the basic modular structures that give rise to the degenerate solutions. To be useful, however, the high-modularity partitions should be sampled in an unbiased and relatively complete way, e.g., by using a Markov chain Monte Carlo algorithm~\cite{massen:doye:2006}. This type of approach may also provide a way to identify overlapping~\cite{sawardecker:sales-pardo:amaral:2009} or hierarchical modules~\cite{sales-pardo:etal:2007}. (That being said, hierarchical structure poses a special problem for modularity, because, strictly speaking, there is no ``correct'' partition of a hierarchy; at best, a good partition will identify the modules at a particular hierarchical level. Separate tools are needed to infer distinct levels.) On the other hand, the difficulty of constructing an unbiased sample of an exponential number of degenerate solutions may prevent these methods from uncovering subtle or large-scale relationships, particularly in larger networks.

Another set of promising techniques try to estimate the statistical significance of a high-modularity partition~\cite{karrer:etal:2008,lancichinetti:etal:2009}, i.e., to answer the question of how much true structure is captured by a particular high-modularity partition. And, techniques based on local methods~\cite{clauset:2005,bagrow:2008}, which do not attempt to partition the entire network, or on random walks over the network~\cite{ziv:middendorf:wiggins:2005,pons:latapy:2006,rosvall:bergstrom:2008,lambiotte:delvenne:barahona:2008}, may provide useful alternatives to modularity maximization, although these may still exhibit degenerate behavior.

A particularly promising class of techniques for identifying modular and hierarchical structures relies on generative models and likelihood functions. Stochastic block models~\cite{rosvall:bergstrom:2007,newman:leicht:2007,hofman:wiggins:2008,airoldi:etal:2008,bickel:chen:2009,wang:wong:1987} and hierarchical block models~\cite{clauset:etal:2008} are attractive because they can allow module densities to vary independently, although their flexibility can come with computational costs and their results can be more difficult to interpret. In some cases, these models can capture overlapping modules~\cite{airoldi:etal:2008}. In general, the likelihood framework presents several opportunities not currently available for modularity-based methods. For instance, by comparing the likelihoods of empirical network data under different structural models, researchers can give statistically principled answers to model selection questions, such as, is this network more hierarchical, more modular, or neither? But, likelihood functions can also exhibit extreme degeneracies and optimization techniques for module identification should likely be treated with caution. To ensure good results, it may be necessary to use a sampling approach~\cite{clauset:etal:2008}.

In closing, we note that the development of objective and accurate methods for identifying modular and hierarchical structures in empirical network data is crucial for many systems-level questions about the structure, function, dynamics, robustness and evolution of many complex systems.  The magnitude of the degeneracy problem, and the dependence of $Q_{\max}$ on the size and number of modules in the network, suggests that modules identified through modularity maximization should be treated with caution in all but the most straightforward cases. That is, if the network is relatively small and contains only a few non-hierarchical and non-overlapping modular structures, the degeneracy problem is less severe and modularity maximization methods are likely to perform well. In other cases, modularity maximization can only provide a rough sketch of some parts of a network's modular organization. We look forward to the innovations that will allow it to reliably yield more precise insights.

\begin{acknowledgments}
The authors thank Luis Amaral, Vincent Blondel, Nathan Eagle, Santo Fortunato, Petter Holme, Brian Karrer, David Kempe, John Lee, Cristopher Moore, Mark Newman and Michel Verleysen for helpful conversations, and Petter Holme and Mikael Huss for sharing network data. \mbox{Y.-A. de M.} thanks Vincent Blondel for supporting a visit to the Santa Fe Institute where this work was conducted. This work was funded in part by the Santa Fe Institute, the National Science Foundation's Research Experience for Undergraduates (REU) Program and a grant from the James S. McDonnell Foundation. 
\end{acknowledgments}

\begin{appendix}
\section{The dependence of $Q_{\max}$ on $n$ for the ring network}
\label{appendix:qmax:ring}
A simple case where it is straightforward to work out the precise dependence of $Q_{\max}$ on network size is the ring network from Section~\ref{section:resolution:limit}.

Consider such a network with $k$ cliques, each composed of exactly $c$ nodes, and where we hold $c$ constant while increasing $n$, i.e., we add more modules to the ring such that $k = n/c$.  If the optimal partition merges $\ell$ adjacent cliques (due to the resolution limit), then it can be shown that the modularity is exactly
\begin{align}
Q_{\max} & =  1 - \frac{1}{\ell} \left(\frac{1}{{c \choose 2} + 1}\right) - \frac{c\, \ell}{n} \enspace , \label{eq:ring:qmax}
\end{align}  
where 
\begin{align}
\ell & =  \left\lfloor \sqrt{ \frac{1}{4} + \frac{n/c}{{c \choose 2} + 1}} - \frac{1}{2} \right\rfloor = O(\sqrt{n}) \enspace . \nonumber
\end{align}  
Thus, as $n \to \infty$ the second and third terms in Eq.~\eqref{eq:ring:qmax} vanish like $O(1/\sqrt{n})$ and $Q_{\max} \to 1$.

As a brief aside, we now connect this result to the large-scale behavior of the ``plateau'' region of the modularity function mentioned in the main text. For concreteness, we define the plateau as the set of partitions with modularity scores within 10\% of $Q_{\max}$.

To begin, we note that the asymptotic result given above implies that the height of the plateau, which is simply the maximum modularity value, increases with $k$. We now characterize the size of the plateau by considering the number of partitions formed by merging connected cliques.  As shown in Section~\ref{section:degeneracies}, there are $2^{k}$ such partitions for the ring network because there are $k$ edges connecting cliques, each of which can be cut or not cut to create a different group.

Let $Q_{1}$ be the modularity score of the intuitive partition, i.e., the one that places each clique in its own group. It can be shown that the ratio $Q_{1}/Q_{\max}$ is a monotonically decreasing function of $k$ whose limit is
\begin{align}
\lim_{k \to \infty} \frac{Q_1}{Q_{\max}} & = 1 - \frac{1}{{c \choose 2} + 1} \enspace .
\end{align}
If the cliques are composed of at least $c=5$ nodes, this ratio is $10/11$ and the intuitive partition $Q_1$ is always somewhere within the plateau region. 

If the optimal partition merges $\ell$ adjacent cliques, then it can be shown that there are at least $2^{k(1- 1/\ell)}$ partitions with no more than $\ell$ cliques in a single module and each of these partitions will be within 10\% of the maximum modularity because $Q_{1}$ bounds their modularity from below.  Since $\ell=O(\sqrt{n})=O(\sqrt{k})$, in the limit of large $k$, the number of partitions depends only on the number of cliques and we have an exponential expansion in the number of partitions in the high-modularity plateau. Thus, as $k$ grows large, both the height and the size of the plateau increase as well, with the latter increasing exponentially.

Although the details would change for a different network structure, in principle, such an exponential expansion in the size of the plateau should be universal. This provides a very broad target for optimization algorithms.

\section{Simulated Annealing}
\label{appendix:sa}
To initialize each simulating annealing (SA) sample run, we start the procedure at a ``random'' partition, in which we first choose a number of communities $k$ and then assign each node to one of these communities with equal probability.

At each step of the algorithm, a modification of the current partition is proposed, e.g., by moving a node from one group to another, by merging two groups or by splitting one group into two. If this modification results in a partition with higher modularity, the current partition is replaced with the proposed one. Otherwise it is replaced with probability ${\rm e}^{-|\Delta Q|/T}$, where $\Delta Q$ is the difference in modularity between the current and proposed partition and $T$ is the \emph{temperature} parameter, which we decrease according to the annealing schedule (see below). If the proposed modification is rejected, we retain the current partition and propose a new modification at the next step. As $T \to 0$, the algorithm is guaranteed to converge to a local optimum in the modularity function.

To implement the algorithm, we must define the set of possible modifications (the \emph{move set}), which determines the local neighborhood of any given partition. Different choices of move set can drastically alter both the convergence time of the algorithm and its ability to escape local optima.  The choice of move set can even affect the kinds of local optima we sample (see below).  For our purposes, it is less important that the algorithm converge on the global optimum than it is to sample a broad section of the modularity function in a relatively unbiased way. Some alternative heuristics for maximizing modularity can also be used to sample the modularity landscape, e.g., the Louvain method~\cite{blondel:etal:2008}, but these often do so with particular biases and thus are not as flexible as simulated annealing for obtaining a clear view of the modularity function's degeneracies (e.g., see Fig.~\ref{fig:metabolic:summaries}).

We employed two simple move sets: (i) single node moves and (ii) a combination of single moves, merges and splits.  A single node move takes a node chosen uniformly at random from the $n$ nodes in the network and either moves it to another group, chosen uniformly at random from the remaining groups, or places it in a new group by itself. (If the chosen node is the only member of its group and it is successfully moved to another existing group, the number of existing groups decreases.) If the current partition has $k$ communities, this move set defines a local neighborhood for any particular partition that is composed of $w_1 = n(k-1) + n = nk$ neighboring partitions. (We note that this move set is similar to the partition modifications used in the Kernighan-Lin heuristic~\cite{kernighan:lin:1970}.)

In the second move set, we also allow merges and splits.  With probability $p_m$ we choose two groups uniformly at random and merge them into a single group.  Alternatively, with probability $p_s$ we choose a group uniformly at random and split it into two subgroups such that the number of edges between them is minimized. (This differs from the ``heat bath'' approach used in Ref.~\cite{guimera:amaral:2005}.) This optimization problem is conventionally called \textsc{Mincut} and we use a standard algorithmic solution for finding the minimum cut weight \cite{stoer:wagner:1997}. This way of choosing a split for a group typically results in a relatively good partition; in contrast, a randomly chosen bipartition would almost surely result in a lower modularity score and thus would almost always be rejected. Finally, with probability $1 - (p_m + p_s)$, we perform a single node move as described above. This move set defines a local neighborhood of size $w_2 = nk + {k \choose 2} + k$, where the first term comes from the single node moves (as above) and the other terms denote the number of merges and splits, respectively.  For a particular network, we choose $p_m$ and $p_s$ so that each individual neighboring partition is proposed with roughly equal probability.

Once the move set is chosen, the convergence of the SA algorithm is determined by the \emph{annealing schedule}, which controls the rate at which the temperature parameter decreases. For simplicity, we use a geometric schedule $T(t) = T_{0\,}r^{t}$, where $T_{0} > 0$ is some initial temperature and $0 < r < 1$ is the common ratio between successive temperatures. For best results, $T_{0}$ and $r$ must typically be tuned to a particular network topology, but so long as they are chosen to allow the SA algorithm sufficient time to explore the partition space, their values do not significantly impact our results.

Each sample run obeys a termination criterion that is derived by bounding the number of failed modifications needed to decide whether the current partition is a local optimum with high probability.  Let $w^{*}$ be the number of moves required to try each of the $w$ possible modifications of the current partition.  It can be shown that
\begin{equation}
\Pr[w^{*} > \beta w \log w] \leq w^{-\beta+1} \enspace . \nonumber
\end{equation}
We choose $\beta$ such that after $\beta nk \log(nk)$ rejected modifications, there is a 95\% chance that there are no modifications that would increase the modularity of the current partition. When this criterion is met, the SA algorithm terminates. The termination criterion is only necessary to improve the running time of the algorithm, particularly toward the end of the annealing schedule when most proposed modifications result in lower modularity scores.

\begin{figure}[t]
\centering
\includegraphics[width=0.95\columnwidth]{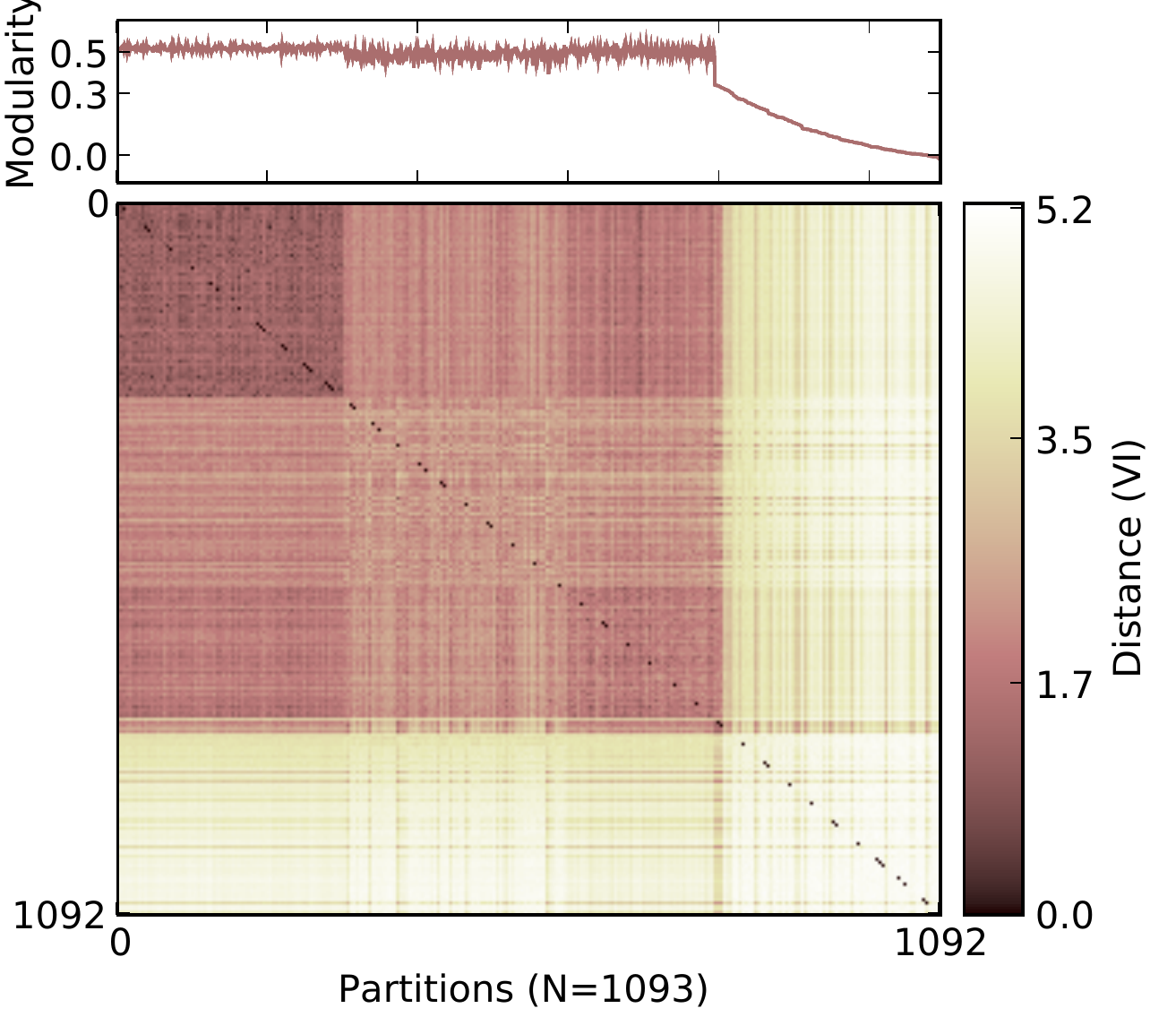}
\caption{(color online) For the metabolic network of {\em Treponema pallidum}, the matrix of pairwise distances calculated from a sample of (i) 301 unique partitions found by the Louvain method, (ii) 292 unique local optima sampled using the single node move set, (iii) 200 unique local optima sampled using the merge-split move set and (iv) unique 300 low-modularity partitions sampled using the single node move set. The inset shows the corresponding modularity score as a function of left-to-right ordering in the matrix. The different sampling methods cause the coarse block structure in the distance matrix.}
\label{fig:algorithm-compare}
\end{figure}

For practical purposes, we made two slight modifications to the SA algorithm described above.  To prevent the algorithm from wasting significant time oscillating between two partitions whose modularity scores are identical, we implement a self-avoiding behavior: in addition to the ordinary acceptance conditions, a proposal is accepted only if it represents a partition that has not previously been visited. This requirement is very unlikely to deny the SA algorithm access to the entire partition space for any but the smallest networks while considerably improving the performance on larger networks.

The second modification concerns the initial partition assignment.  Instead of choosing an initial value for $k$ uniformly from the set $\{1, \ldots, n \}$, we first select a value $k_{\max} \leq n$ and  choose $k$ uniformly from $\{1,\ldots k_{\max}\}$.  For large networks, this prevents the algorithm from spending considerable amounts of time reducing the number of groups from $O(n)$ down to a more appropriate value, which mainly impacts the running time of the algorithm.

\subsection*{On the choice of move set and alternative algorithms}
There are any number of alternative move sets we could have employed and we intentionally considered only the two described above. This choice is motivated partly by convention, as previous SA algorithms for modularity maximization~\cite{guimera:amaral:2005} have employed similar move sets, and partly on theoretical grounds, as the single node move set constitutes the most natural minimal changes to a partition while merge-split moves constitute the most natural higher-order or large-scale change to a partition in a modular network. Thus, our choices are principled, but they are not guaranteed to be optimal. It is theoretically possible that there exists a move set, i.e., a way of defining which partitions are ``local'' to each other, such that the degeneracy problem we describe largely disappears and the modularity function seen by this algorithm exhibits a clear and easy-to-find global optimum. However, the NP-hardness result of Brandes et al.~\cite{brandes:etal:2008} implies that, in general, there can be no such ideal move set for modularity maximization, i.e., one that allows us to efficiently find the global optimum, unless P=NP~\cite{garey:johnson:1979}.

Alternative heuristics for optimizing the modularity function implicitly choose different move sets than the ones described above. Thus, different algorithms will ``see'' different versions of the modularity function and they may sample or target distinct high-modularity regions of the function. To test whether our results from SA are specific to the SA framework and our selected move sets, we briefly consider whether the partitions sampled by a very different heuristic---Blondel et al.'s Louvain method~\cite{blondel:etal:2008}, which builds a high-modularity partition by recursively agglomerating groups of connected nodes or modules until a high modularity is achieved---exhibit similar behavior or overlap with those sampled by the SA approaches.

Using the {\em Treponema pallidum} metabolic network as a realistic test case, we sample several hundred high-modularity partitions using the Louvain method, several hundred using the single node move set, and several hundred using the move-split move set. For comparison, we include several hundred low-modularity partitions from the single node move set (sampled early in the SA). Figure~\ref{fig:algorithm-compare} shows the resulting matrix of pairwise distances for these partitions (measured by their variation of information; see Section~\ref{appendix:distances} below).

Most notably, we see that there is very little overlap between the high-modularity partitions sampled by the three heuristics, suggesting that different move sets (and thus different algorithms) do indeed sample distinct parts of the modularity function. In fact, the partitions sampled by the two SA move sets overlap very little. The fact that these sampled regions are distinct but still exhibit very high modularities (inset) reinforces the fact that the degeneracy phenomenon is ubiquitous and suggests that other approaches are likely to face similar issues.

\section{The Distance Between Partitions}
\label{appendix:distances}
We quantify the differences between partitions using one popular notion of partition ``distance''  called the variation of information (${\rm VI}$), which was introduced by Meil\u{a}~\cite{meila:2005}. This measure satisfies the standard axioms for a distance metric and thus preserves many of the intuitive properties we expect from a distance measure. Further, it does not rely on finding a maximally overlapping alignment of the partitions, which makes it fast to calculate. For a thorough discussion of other notions of distance between partitions, and of the advantages of the ${\rm VI}$ measure, see Ref.~\cite{karrer:etal:2008}.

The ${\rm VI}$ allows us to quantitatively test the hypothesis that suboptimal high-modularity partitions disagree with the optimal partition mainly in small or trivial ways, which would correspond to very small ${\rm VI}$ values (close to $0$), e.g., Fig.~\ref{fig:metabolic:coarse}. It also allows us to construct low-dimensional visualizations of the sampled modularity function.

\begin{figure}[t]
\centering
\includegraphics[scale=0.55]{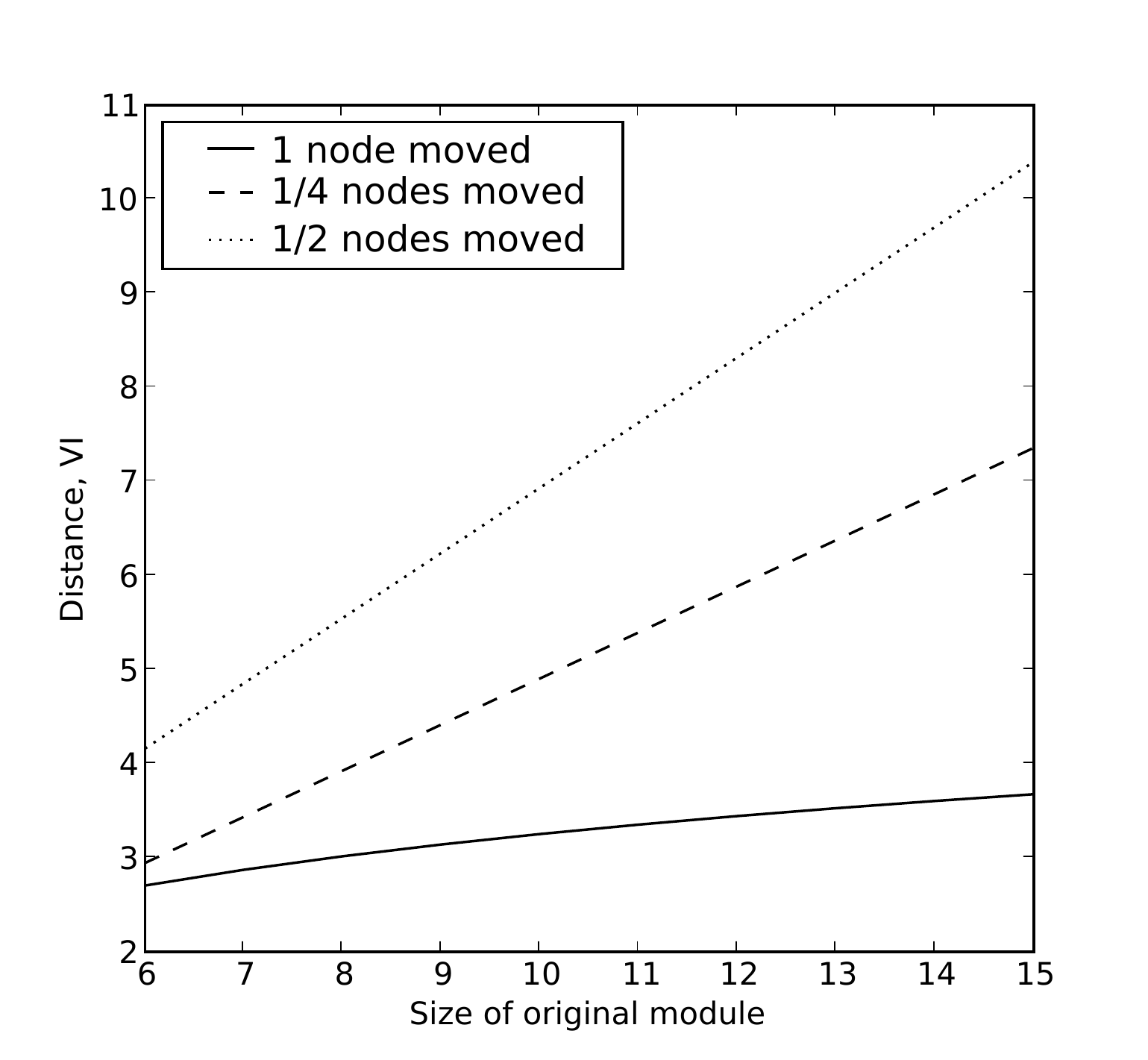}
\caption{The variation of information (${\rm VI}$), as a function of the size of the original module, when we move a single node into a new group, move 1/4 of the original nodes into a new group, or move 1/2 of the original nodes into a new group. In all cases, the ${\rm VI}$ increases monotonically, but with a slope that depends on the fraction of the original module being moved.} 
\label{fig:vi-demo}
\end{figure}

Given partitions $C$ and $C'$, their ${\rm VI}$ is defined as
\begin{align}
{\rm VI}(C,C') & = H(C) + H(C') - 2I(C;C') \\
 &= H(C,C') - I(C;C') \enspace , \label{eq:vi}
\end{align}
where $H(.)$ is the entropy function and $I(.;.)$ is the mutual information function. Using the definitions
\begin{align}
H(C,C') & = -\sum_{i,j}p(i,j)\log p(i,j) \nonumber \\
& = -\sum_{i,j} \frac{n_{i,j}}{n}\log\left(\frac{n_{i,j}}{n} \right) \label{eq:joint:entropy} \\
I(C;C') & = \sum_{i,j}p(i,j)\log\left( \frac{p(i,j)}{p(i)p(j)} \right) \nonumber \\
&= \sum_{i,j}\frac{n_{i,j}}{n}\log\left(\frac{n_{i,j}\,n}{n_{i}n_{j}} \right)  \label{eq:mutual:information}
\end{align}
we can further simplify Eq.~\eqref{eq:vi} to an expression that depends only on counts:
\begin{equation}
{\rm VI}(C,C') = - \frac{1}{n} \sum_{i,j} n_{i,j} \log\left( \frac{n_{i,j}^2}{n_i n_j} \right) \enspace ,
\end{equation}
where $n_i$ is the number of nodes in group $i$ in $C$, $n_j$ is the number of nodes in group $j$ in $C'$ and $n_{i,j}$ is the total number of nodes in group $i$ in $C$ and in group $j$ in $C'$. Two partitions of the network are the same if and only if ${\rm VI}(C,C') = 0$ and the maximum possible ${\rm VI}$ is given by $\log n$ where $n$ is the number of nodes in the network.

\subsection*{Two example calculations using ${\rm VI}$}
To give the reader a more intuitive feeling for how ${\rm VI}$ behaves, we briefly calculate a few distances using the mis-merged partitions we encountered in the main text.

First, consider a partition, with $k$ modules, in which one module has $g$ nodes. If we move $h$ nodes from this module into a new group, the distance between the original and the new partition is
\begin{align}
{\rm VI}(C,C') & = \frac{1}{n} \left[ g \log g - (g - h) \log ( g- h) - h \log h \right] \enspace ,
\end{align}
which obtains its maximum of $(g/n) \log 2$ when $h = g/2$.  Fig.~\ref{fig:vi-demo} shows the functional dependence of the ${\rm VI}$ for several choices of $g$ and $h$. Most notably, under ${\rm VI}$, partitions that differ by a merge of two groups or a split of one group are more distant than those that differ only by a few displaced nodes. From the discussion in the main text, partitions that differ by merges and splits are precisely the kind we expect to find among the high-modularity but suboptimal partitions. 

For a second example, consider the split and merge operation discussed for a hierarchical network in the main text, where the modules $i=\{a,b\}$ and $j=\{c,d\}$ are both of size $g$ and their submodules contain $g/2$ nodes each. The alternate partition $i'=\{a,c\}$ and $j'=\{b,d\}$ has a distance 
\begin{equation}
{\rm VI}(C,C') = 4 (g/n) \log 2 \enspace ,
\end{equation}
from the original one. Thus, this split and merge operation produces a partition that is four times the distance from the original partition as one obtained by a single bisection of one group (the previous example).

Finally, we note that the ${\rm VI}$ notion of distance is not without its weaknesses. The most significant of these is its unintuitive scale. Further, the maximum ${\rm VI}$ scales with the number of nodes or the number of modules in the partition and thus we cannot reliably compare ${\rm VI}$ distances between networks with different sizes or number of modules. Thus, our results here and in the main text rely only on relative distances for partitions of the same network and not on any particular numerical value.

\subsection*{Alternative distance measures}

Although the variation of information is a satisfactory partition distance measure for our needs, we would like to ensure that our main results (e.g., the ruggedness of the high modularity plateau and the large-scale structural disagreements between the high-modularity partitions) do not depend sensitively on the distance measure used. All that we technically require is a distance measure that satisfies the standard metric axioms. Another possible choice is provided by the \emph{Jaccard distance} $J$, which is defined as
\begin{equation}
J(C,C') = \frac{a_{01} + a_{10}}{{n \choose 2}} \enspace ,
\end{equation}
where $a_{01}$ is the number of pairs of nodes that are in the same module in $C$ but different modules in $C'$ and vice versa for $a_{10}$.   

\begin{figure}[t]
\begin{center}
\includegraphics[scale=0.235]{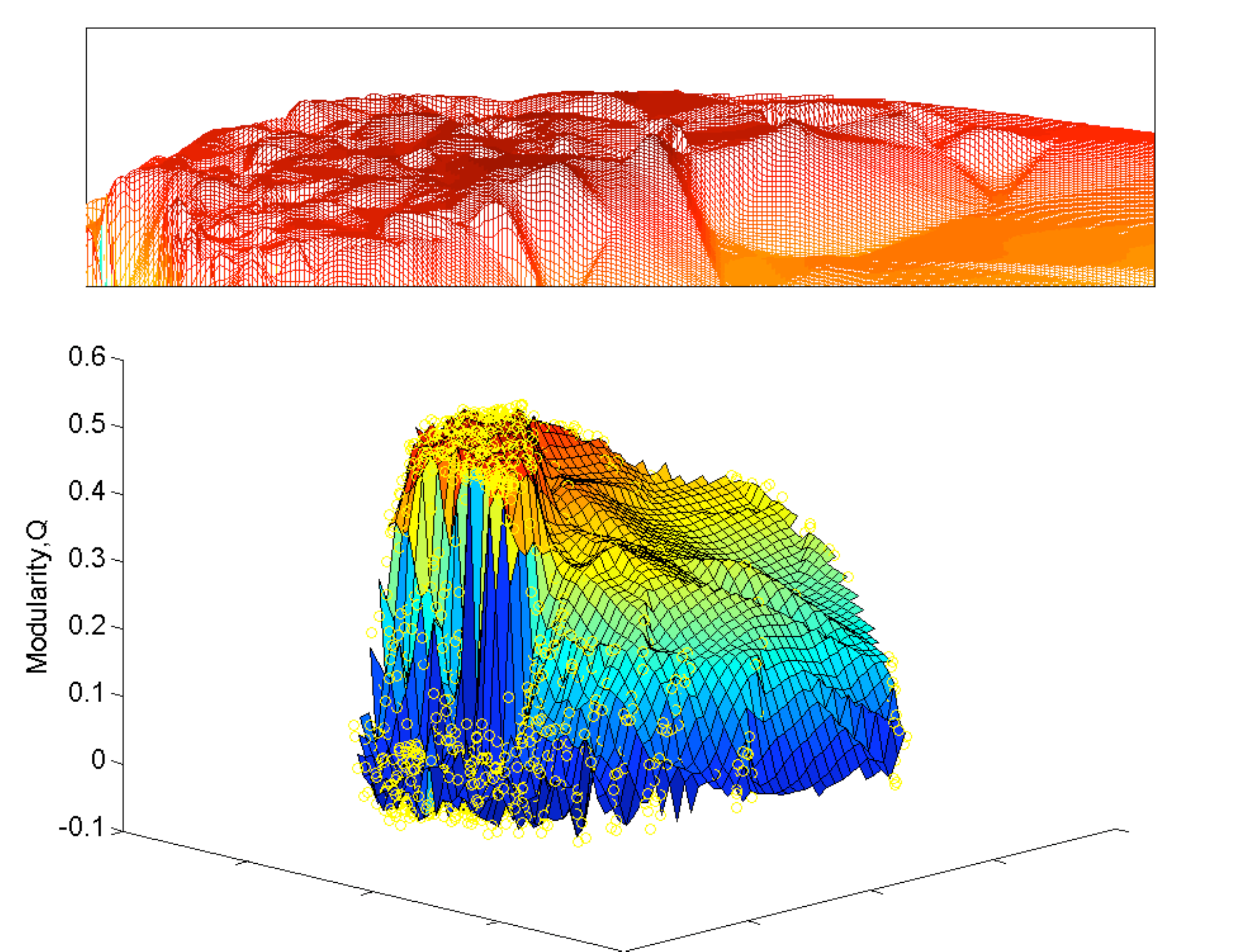} 
\caption{(color online) The modularity function for the metabolic network of the spirochaete {\em Treponema pallidum} reconstructed using the Jaccard distance, which shows the same qualitative structure that we observed using the variation of information. The inset shows the rugged high-modularity region. This suggests that our results do not depend sensitively on the choice of distance metric.}
\label{fig:metabolic:jaccard}
\end{center}
\end{figure}

Figure~\ref{fig:metabolic:jaccard} shows the reconstructed modularity function for the \emph{T. pallidum} metabolic network using the Jaccard distance in place of the variation of information. Although the precise shape of the modularity function is different, we observe the same qualitative behavior present in the VI landscape: a rugged high-modularity plateau surrounded by a sea of lower modularity partitions. Furthermore, the Jaccard distance yields similar results to the variation of information when conducting the coarse-graining analysis outlined in Section~\ref{section:realworld:networks}. This suggests that our results do indeed capture real properties of the modularity landscapes of these networks and do not depend on our choice of distance measure.

\section{Curvilinear Component Analysis}
\label{appendix:embedding}
In principle, a matrix of pairwise ${\rm VI}$ distances for partitions of a network (like the one shown in Fig.~\ref{fig:algorithm-compare}) contains all the information necessary to understand the structure of the modularity function. However, the non-Euclidean nature of the partition space makes this information difficult to interpret. Thus, we use an embedding algorithm to project the distance matrix onto a two-dimensional Euclidean landscape. The modularity scores of each partition provide a third dimension.

The projection from the original space (hereby referred to as the \emph{data space}) to the 2D landscape (known as the \emph{latent space}) can be phrased as an optimization problem: we seek an assignment of partitions to positions in the latent space that preserves the original pairwise distances as much as possible. The quality of any particular assignment is conventionally characterized by a stress function, which measures the errors in the projected distances.

We use the curvilinear component analysis (CCA) algorithm~\cite{lee:verleysen:2007}, which preserves local distances at the expense of some amount of distortion at larger distances. (Other suitable embedding algorithms exist, e.g., Sammon's non-linear mapping~\cite{sammon:1969}, but these often have concave error functions and are thus not guaranteed to converge.) Given a set of distances $d_D(x,y)$ in the data space, we wish to assign distances $d_L(x,y)$ in the latent space so as to minimize the stress function:
\begin{align}
E_{cca} & = \frac{1}{2} \sum_{x,y}\left[d_D(x,y)-d_L(x,y)\right]^2 F_{\lambda}(d_L(x,y)) \enspace ,
\end{align}
where $F_{\lambda}$ is a weight function. Here, we take $F_{\lambda}$ to be a linear combination of Heaviside step functions chosen to produce a decreasing function with a null first derivative nearly everywhere (for details, see Ref.~\cite{demartines:herault:1995}).  This choice tends to conserve shorter distances while occasionally producing ``tears" for large distances. The stress function is then minimized using the optimization procedure designed by Demartines and Herault~\cite{demartines:herault:1995}.

\begin{figure}[t]
\centering
\includegraphics[scale=0.45]{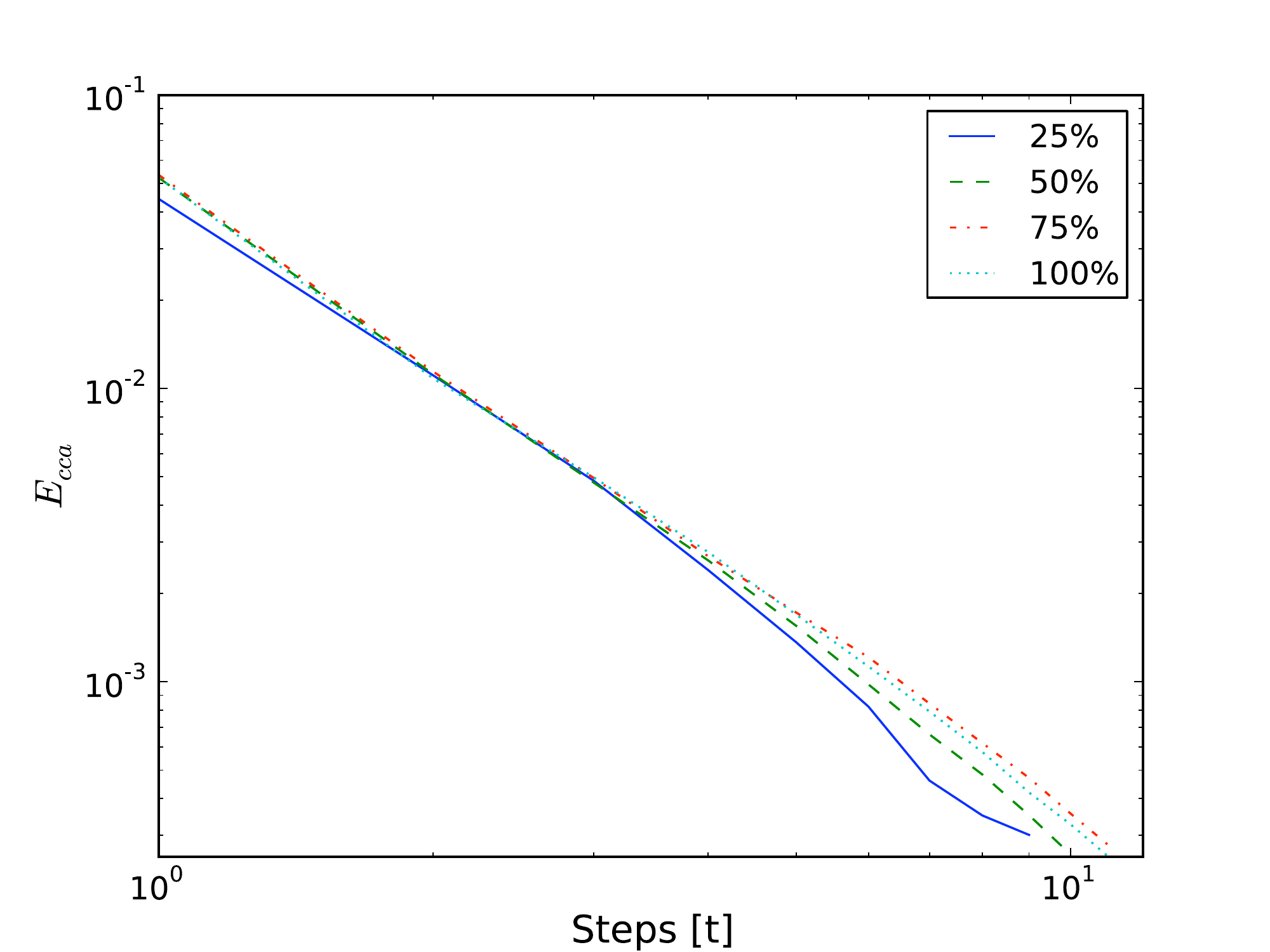}
\caption{(color online) The error rate of the embedding $E_{cca}$ as a function of the number of steps $t$ in the optimization algorithm for 25\%, 50\%, 75\% and 100\% of the 997 samples for the ring network (Fig.~2), normalized by sample size. The $O(t^{-3})$ decay in the error rate shows that the CCA algorithm is robust to the number of samples and provides highly accurate a embedding of the original distances.}
\label{fig:errorCCANorm}
\end{figure}

In order to generate a relatively unbiased sampling of the modularity function from a large set of independent SA runs, i.e., to ensure that we sample both high and low modularity partitions and that our samples are relatively independent of each other, we do the following. A quarter of our sampled partitions are obtained by choosing the local optimum found when the run terminates.  Each remaining partition is chosen by running the SA algorithm to its $t$th step, where $t$ is drawn iid from a geometric distribution. By drawing only one partition from each run, and combining results from a large number of independent runs, we obtain a relatively even sampling of the high-modularity region of the modularity function.

Notably, this procedure does not produce an unbiased sample, which could be obtained using a Markov chain Monte Carlo technique~\cite{massen:doye:2006}. However, our goal is not a fully unbiased sample of partitions; rather, we seek a sufficiently even and unbiased sample of the high-modularity partitions that we can study the question of the modularity function's degeneracies and get a realistic reconstruction of this region of the modularity function. By biasing our sample in favor of high-modularity partitions, but sampling them independently, we can achieve that goal. The partitions with intermediate modularity values are included to ensure some coverage of mid- and low-modularity regions.

\begin{figure}[t]
\centering
\includegraphics[scale=0.50]{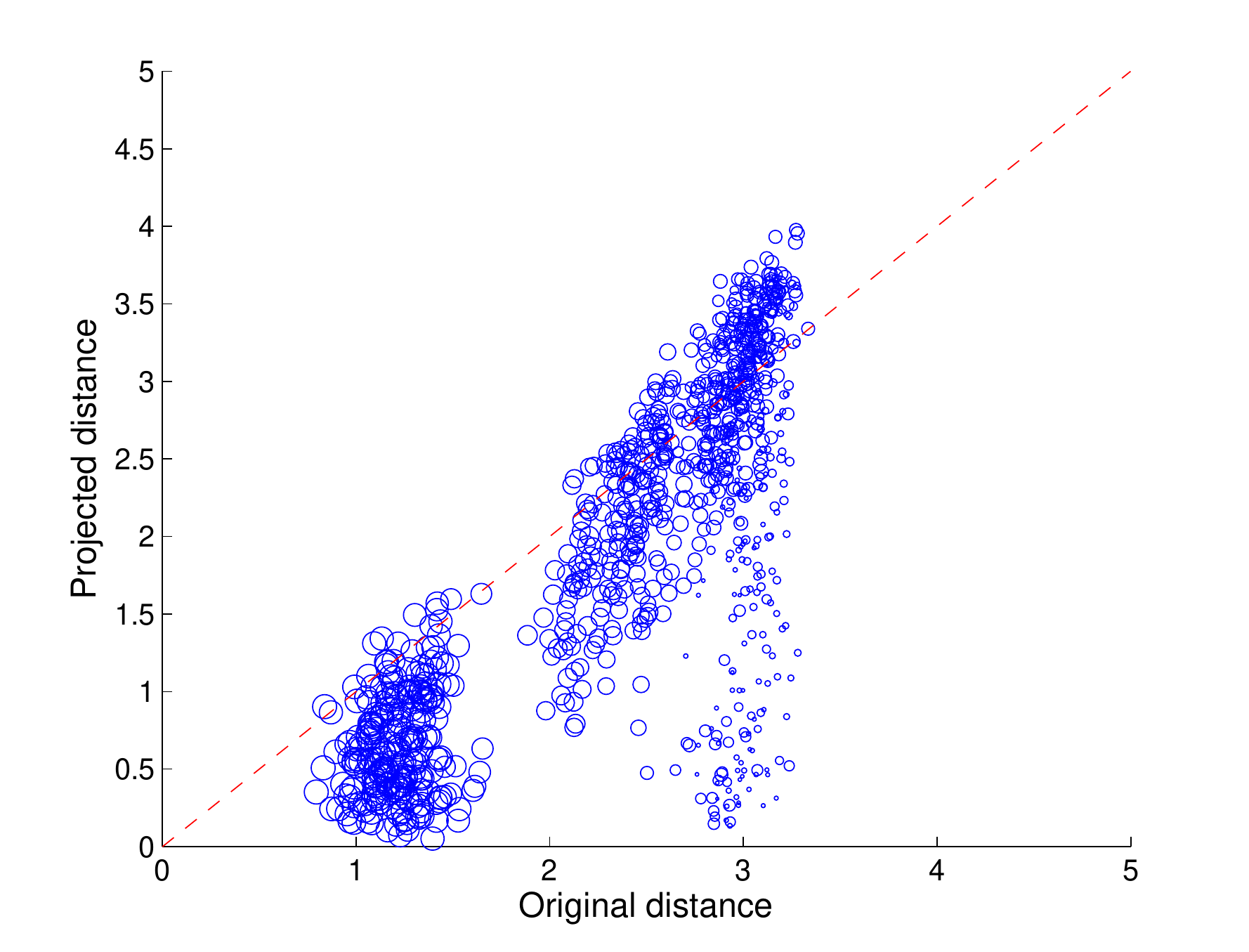}
\caption{(color online) Latent space distances as a function of data space distances for a sample of 997 partitions of the ring network. Point sizes are weighted by the average modularity of the two partitions: larger circles represent distances between two high-modularity partitions whereas small circles correspond to distances between low modularity partitions.}
\label{fig:ShepardDiagRing25}
\end{figure}

We validate the results of our embeddings in three ways.  First, we test whether the qualitative structure of the embedded functions depends on the number of samples used.  Using the ring network, we subsampled the 997 partitions used to construct Fig.~\ref{fig:ring:partitions} at the 25\%, 50\% and 75\%  levels. Adding more samples should never decrease the ruggedness in the high-modularity region, but if the landscape changes significantly, it could indicate a problem with the embedding. Comparing the results, we find that the qualitative structure of the four landscapes---including the rugged structure of the plateau region (Fig.~\ref{fig:ring:partitions}, inset)---is independent of the subsampling rate, suggesting that our full sample is more than adequate to give an accurate representation of the modularity function's structure.

Second, we verify that the decrease in the stress function $E_{cca}$ is well behaved as the number of optimization steps increases, i.e., we see no evidence for pathological behavior in the embedding procedure. For all four of the subsampling levels described above, we find that the error decays roughly as $O(t^{-3})$ in the number of optimization steps $t$ (Fig.~\ref{fig:errorCCANorm}) and the mean final error is roughly $10^{-4}$. Since the mean distance between points is of order 1, this error rate implies that the embedding is quite accurate.

Finally, we test whether our choice of $F_{\lambda}$ conserves the local structure of the modularity function.  We test this by means of a \emph{Shepard diagram}~\cite{kruskal:shepard:1974}, which plots a random sample of the distances in the data space against the corresponding distances in the latent space. A Shepard diagram for the embedded ring network is shown in Fig.~\ref{fig:ShepardDiagRing25}.  We note that deviations from the diagonal occur primarily at larger distances and that the local structure (bottom left of the figure) is generally well preserved.  Even for those points where distance is not preserved, the algorithm errs on the side of assigning smaller distances, which would only tend to make the landscape appear less rugged, i.e., more smooth, than it truly is.

\section{Hierarchical Random Graphs}
\label{appendix:hrg}

The hierarchical random graph (HRG) model, recently introduced by Clauset, Moore and Newman~\cite{clauset:etal:2008}, provides a simple but realistic way to generate networks with hierarchical structure. However, the full HRG model is too flexible for our purposes. Instead, we employ a simplified version that fixes the hierarchical structure and the way the internal probability values vary.

\begin{figure}
\centering
\includegraphics[scale=0.645]{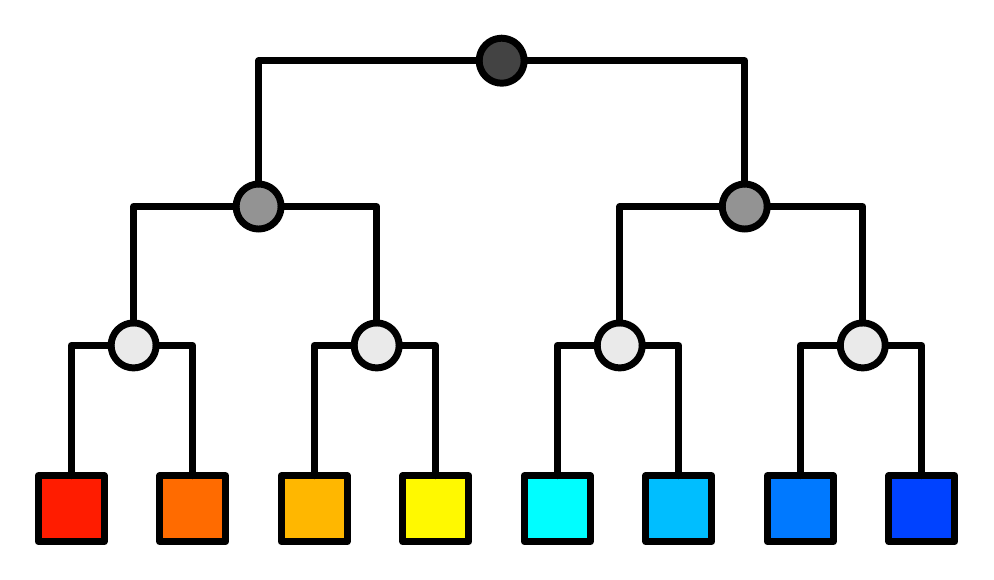} 
\caption{(color online) An example of our simplified hierarchical random graph (HRG) model, with 8 nodes and 4 levels (including the leaves), in which the nodes are organized into a balanced binary tree and the internal probabilities increase as you move from the root toward the leaves.}
\label{fig:tree}
\end{figure}

Under our simplified model, we arrange $n=2^{d_{max}}$ nodes into groups according to a balanced binary tree structure with $d_{max}+1$ levels (Fig.~\ref{fig:tree}). We assign edges between nodes by letting the internal probability values $p_{r}$ increase with their distance from the root of the tree. This regularity gives the network an assortative structure, in which modules become more dense as we move down in the dendrogram. Mathematically, we say that if the lowest common ancestor of two nodes is at level $d$ in the tree, they are connected with probability 
\begin{align}
p_{r}(d) & = 2^{d + 1 - d_{max}} \enspace .
\end{align}

\subsection*{The optimal partition of a hierarchical network}
For this simplified HRG model, we now derive an estimate of the level of the hierarchy whose group structure yields the optimal partition. For convenience, we take a mean-field approach and consider the average modularity $\langle Q \rangle$ of an ensemble of instances drawn from the model. In this case, the modularity function takes the form
\begin{equation}
\langle Q \rangle \approx \sum_{i=1}^{k} \left[ \frac{\langle e_i \rangle}{\langle m \rangle} - \left( \frac{\langle d_i \rangle}{2 \langle m \rangle} \right) ^2 \right] \enspace . 
\label{eq:mean:modularity}
\end{equation} 
Because of the symmetry of the binary tree, the optimal partition must consist of groups of the same size. That is, to find the maximum modularity partition, we must simply find the level $d^{*}$ in the hierarchy that maximizes Eq.~\eqref{eq:mean:modularity}. Accounting for the regular way the group structure changes with the height $d$ from the bottom of the tree, this implies that Eq.~\eqref{eq:mean:modularity} simplifies to
\begin{equation}
\langle Q \rangle =  1 - \frac{d}{d_{max}} - 2^{-d} \enspace .
\end{equation} 
If we treat $d$ as a continuous variable, we find that $\langle Q \rangle$ is maximized when we cut the dendrogram at 
\begin{equation}
d^{*} = \log_2 ( n \ln 2 ) \enspace .
\end{equation}

Thus, this balanced and assortative hierarchical network has a particular behavior with respect to its resolution limit~\cite{fortunato:barthelemy:2007}, i.e., the resolution limit causes the optimal level to move up in the hierarchy as the network grows. The resolution limit always implies that the optimal partition is composed of agglomerations of smaller modules, but in this hierarchical network, these agglomerations are simply composed of modules from lower down in the hierarchy. This analysis, however, says nothing about the degeneracies that characterize the modularity function in the local neighborhood of the optimal partition, which we discussed in Section~\ref{section:degeneracies}.

\begin{figure*}[t]
\begin{center}
\begin{tabular}{cc}
\includegraphics[scale=0.235]{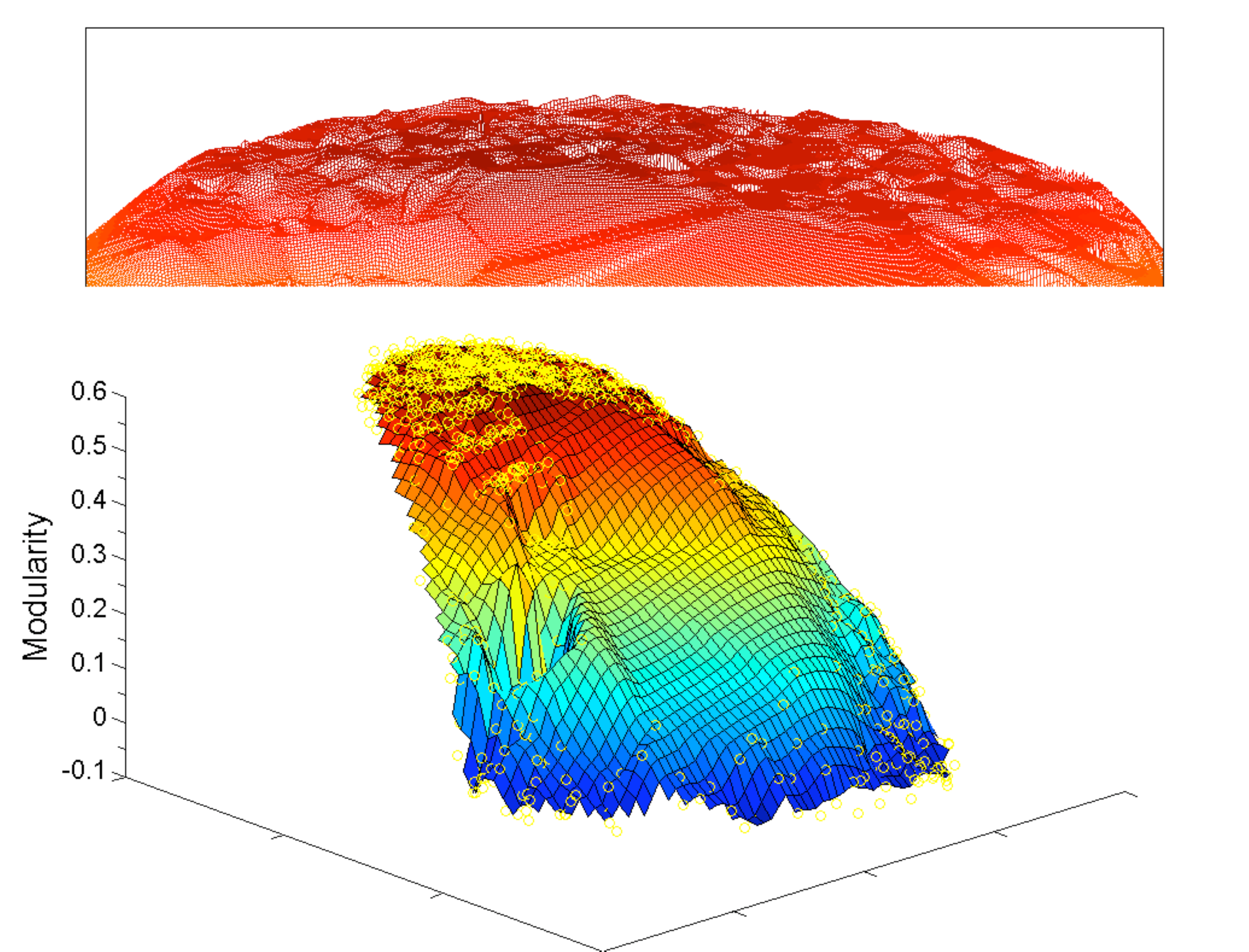} &
\includegraphics[scale=0.235]{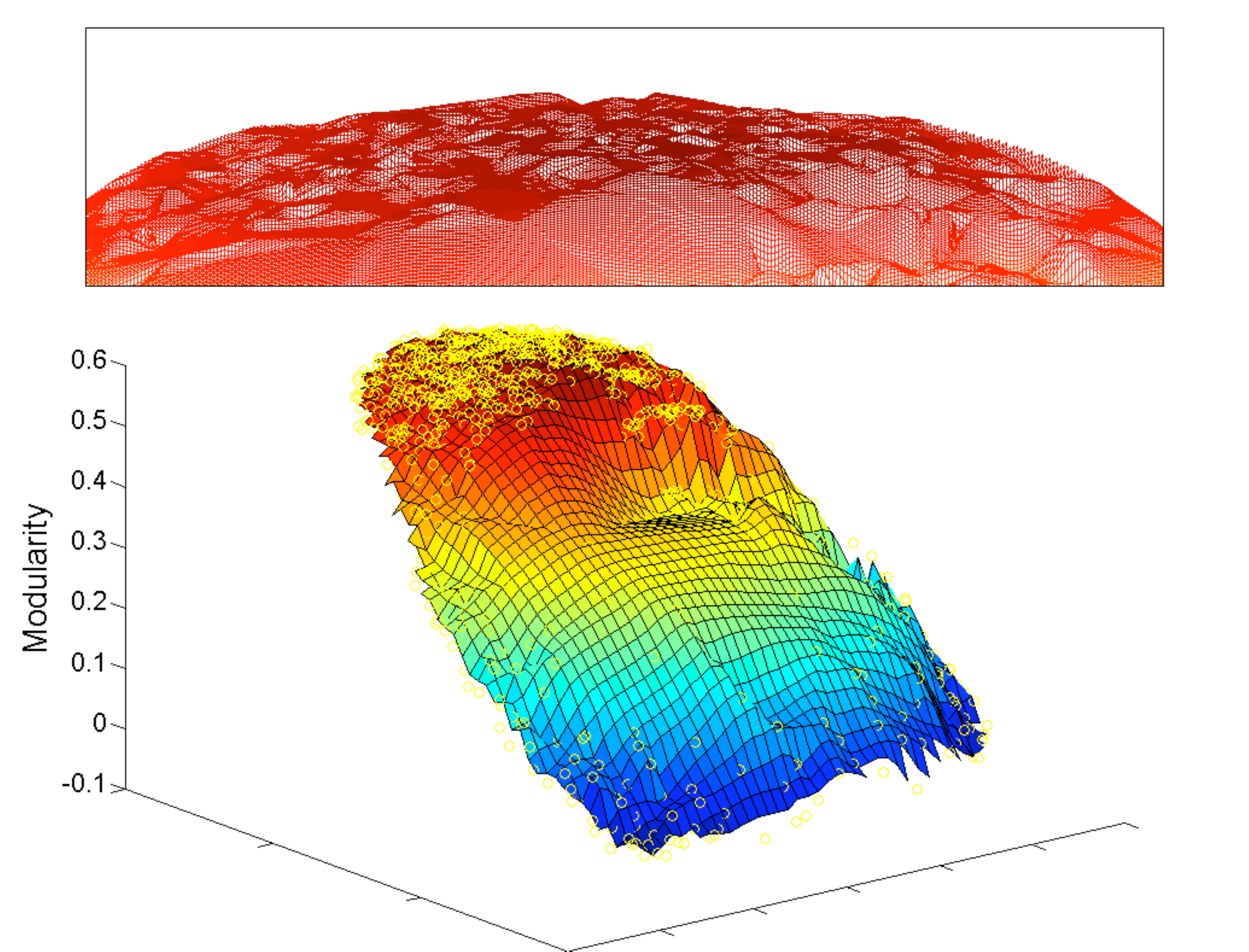}
\end{tabular}
\caption{(color online) Reconstructed modularity functions for the metabolic networks of the mycoplasmatales {\em Mycoplasma pneumoniae} (upper; largest component; $n=354$ and $m=856$) using $1199$ sampled partitions and {\em Ureaplasma parvum} (lower; largest component; $n=300$ and $m=712$) using $1199$ sampled partitions, each showing a large amount of degeneracy among the high-modularity partitions (insets).}
\label{fig:metabolics:more}
\end{center}
\end{figure*}

\begin{figure*}[t]
\begin{center}
\begin{tabular}{cc}
\includegraphics[scale=0.535]{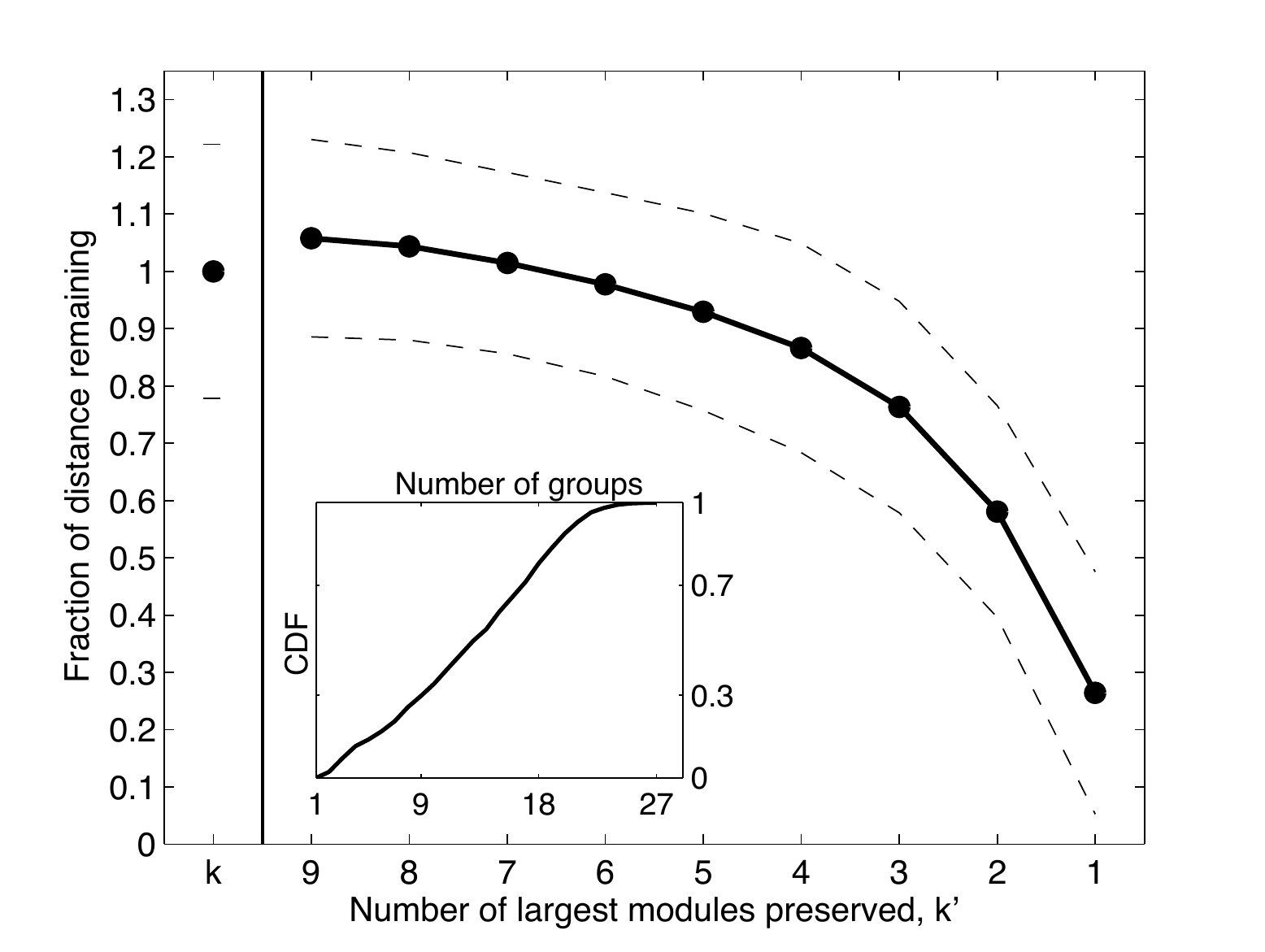} &
\includegraphics[scale=0.535]{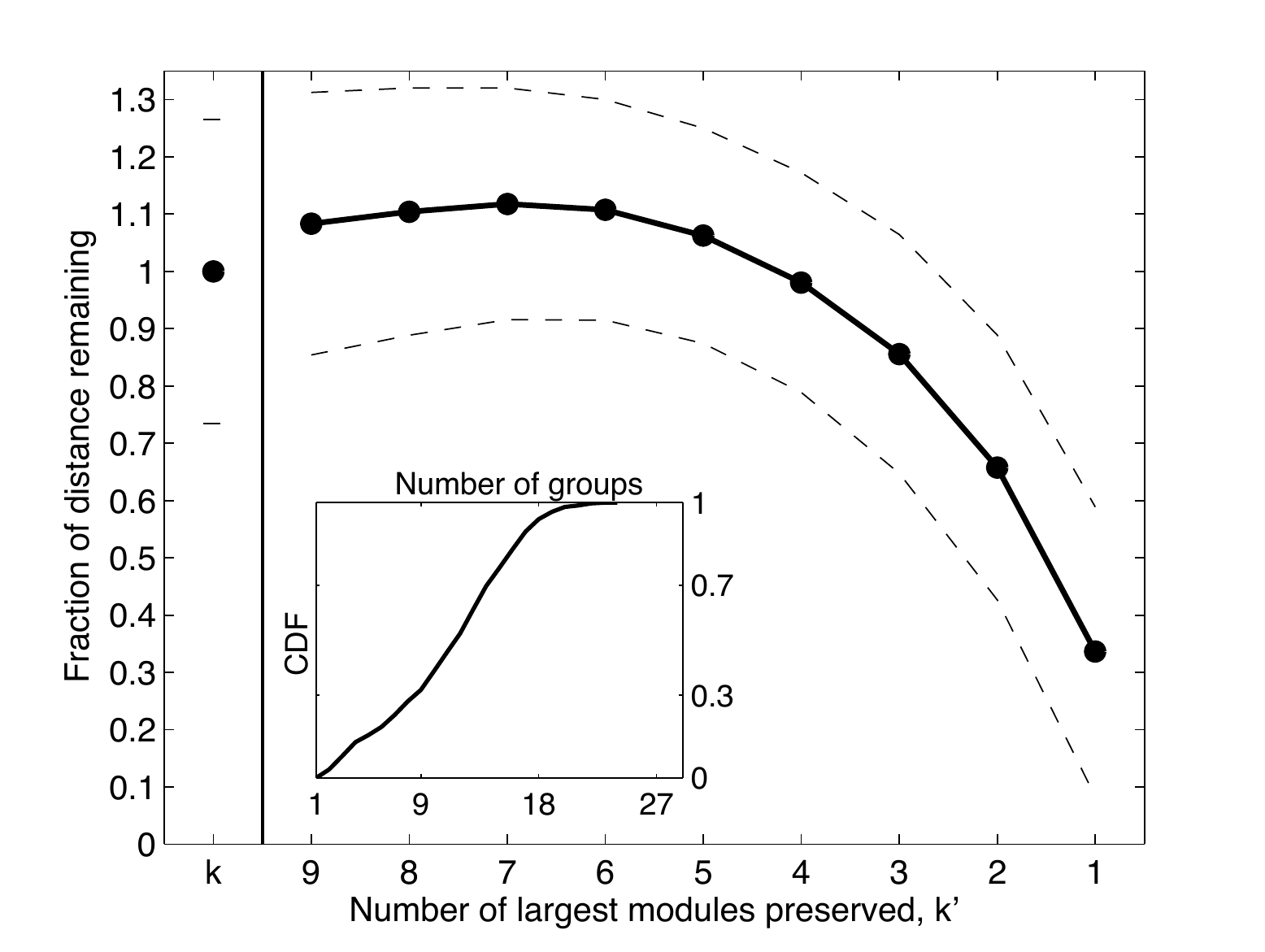} 
\end{tabular}
\caption{For {\em M. pneumoniae} (left) and {\em U. parvum} (right), the fraction of the mean pairwise variation of information (distance) between the sampled high-modularity partitions  that remains when all but the $k'$ largest groups in each partition are merged into a single group. (Dashed lines indicate one standard deviation; insets show the cumulative distributions of the number of partitions with $k$ groups.) Notably, as for the {\em T. pallidum} network discussed in the main text (Fig.~5), the distance distributions change very little when all but the largest few groups in each partition are combined, indicating that most of the distance between partitions is driven by significant differences in the composition of the largest few groups. }
\label{fig:metabolics:sensitivity}
\end{center}
\end{figure*}

\section{Additional Results for Metabolic Networks}
\label{appendix:metabolic:networks}
Figure~\ref{fig:metabolics:more} shows the reconstructed modularity functions for two additional metabolic networks, for the mycoplasmatales {\em Mycoplasma pneumoniae} and {\em Ureaplasma parvum} (3 ATCC 700970)~\cite{huss:holme:2007}.

Figure~\ref{fig:metabolics:sensitivity} shows the corresponding coarsening analyses (analogous to Fig.~\ref{fig:metabolic:coarse}), which confirms that the behavior of the {\em T. pallidum} described in Section~\ref{section:realworld:networks} also holds for these other two networks. That is, for these other networks, we also find significant variation in the composition of the largest few identified modules across the high-modularity partitions, implying that the degeneracies in the modularity function extend beyond simple rearrangements of the smallest modules. Each inset shows the cumulative distribution of the number of groups in the sampled partition. Notably, for all three networks, the fraction of partitions with $k\leq9$ groups is not large enough to explain the persistence of non-trivial distances when all but the largest groups are merged.

Note: the fraction of the original mean pairwise distance ${\rm VI}$ shown in Fig.~\ref{fig:metabolic:coarse} and Fig.~\ref{fig:metabolics:sensitivity} is not guaranteed to decrease monotonically with $k'$. To see why, consider the pair-wise geographic distances between all the cities in California and New York City. The average pairwise distance is composed of two parts: the average pairwise distance between Californian cities and the average distance from each Californian city to New York City. If there are very many Californian cities in our calculation, the overall pairwise average will tend to be dominated by the former term, which has $O(n^{2})$ elements, rather than the latter, which has only $O(n)$ elements. As we merge cities within California, the size of the first term decreases and the average distance becomes more representative of the distance between California and NYC.

\end{appendix}


\end{document}